\documentclass[aps,natbib,nofootinbib]{revtex4-1}

\usepackage{graphicx,amsmath,amsfonts,amssymb,dcolumn,amsthm}
\usepackage{hyperref}
\usepackage{latexsym}
\usepackage{fontenc}
\usepackage[font=small,labelfont=bf,justification=justified]{caption}
\usepackage{color}

\begin{document}

\title{Static and spherically symmetric solutions in a scenario with quadratic curvature contribution}

\author{Fernanda A. Silveira}
\email{fernanda.alvarim@gmail.com}
\affiliation{UFF -- Universidade Federal Fluminense, Instituto de F\' isica}
\altaffiliation{Campus da Praia Vermelha, Avenida General Milton Tavares de Souza s/n, 24210-346, Niter\'oi, RJ, Brazil.}
\affiliation{UERJ -- Universidade Estadual do Rio de Janeiro, Departamento de F\'isica Te\'orica}
\altaffiliation{Rua S\~ao Francisco Xavier 524, 20550-013, Maracan\~a, Rio de Janeiro, RJ, Brasil.}
\author{Rodrigo F. Sobreiro}
\email{rodrigo$\_$sobreiro@id.uff.br}
\affiliation{UFF -- Universidade Federal Fluminense, Instituto de F\' isica}
\altaffiliation{Campus da Praia Vermelha, Avenida General Milton Tavares de Souza s/n, 24210-346, Niter\'oi, RJ, Brazil.}
\author{Anderson A. Tomaz}
\email{tomaz@cbpf.br}
\affiliation{UFF -- Universidade Federal Fluminense, Instituto de F\' isica}
\altaffiliation{Campus da Praia Vermelha, Avenida General Milton Tavares de Souza s/n, 24210-346, Niter\'oi, RJ, Brazil.}
\affiliation{CBPF -- Centro Brasileiro de Pesquisas F\'isicas}
\altaffiliation{Rua Dr. Xavier Sigaud, 150 , Urca, 22290-180, Rio de Janeiro, RJ, Brazil}


\begin{abstract}

In this work we investigate analytic static and spherically symmetric solutions of a generalized theory of gravity in the Einstein-Cartan formalism. The main goal consists in analyzing the behaviour of the solutions under the influence of a quadratic curvature term in the presence of cosmological constant and no torsion. In the first incursion we found an exact de Sitter-like solution. This solution is obtained by imposing vanishing torsion in the field equations. On the other hand, by imposing vanishing torsion directly in the action, we are able to find a perturbative solution around the Schwarzschild-de Sitter usual solution. We briefly discuss classical singularities for each solution and the event and cosmological horizons. A primer discussion on the thermodynamics of the geometrical solutions is also addressed.

\end{abstract}

\pacs{11.15.-q, 04.60.-m, 04.50.Kd}

\maketitle

\section{Introduction}\label{INTRO}

In the present work we consider a generalization of general relativity (GR) \cite{Einstein:1916vd} with cosmological constant in the Einstein-Cartan (EC) formalism \cite{Utiyama:1956sy,Kibble:1961ba,Sciama:1964wt,Zanelli:2005sa}, \emph{i.e.}, where the fundamental variables are the vierbein and the spin connection instead the metric tensor and the affine-connection as in the Palatini formalism \cite{Bergmann:1980wt,Ferraris1982}. In this generalization, the usual Einstein-Hilbert action \cite{DeSabbata:1986sv} is supplemented by a quadratic curvature term and a quadratic torsion term. 

Specifically, we study vacuum static and spherically symmetric solutions of this model by considering the case of vanishing torsion. Because we are considering the EC formalism, the action provides two field equations, one for the vierbein and another for the spin connection. First, we show that a de Sitter spacetime is an exact vacuum solution of these equations. This is a non-trivial result since the system of equations is over-determined for vanishing torsion. After that, we obtain a perturbative solution around the Schwarzschild-de Sitter solution by neglecting the spin connection equation, which is equivalent to impose vanishing torsion at the action level. It is important to be clear that, to obtain such solution, the curvature squared term is treated as perturbation around the usual Einstein term with cosmological constant. Hence, it is the generalized Einstein equation which is perturbed instead a perturbed solution around a fixed background. 

Static and spherically symmetric solutions in alternative gravity models is a recurrent subject of investigation \cite{Stelle:1977ry,Seifert:2007fr,Sebastiani:2010kv,Lu:2015psa,Nojiri:2013su,Cognola:2015uva,Cognola:2015wqa}. In particular, K.~Stelle investigated this subject in a quite general higher derivative scenario in the metric formalism \cite{Stelle:1977ry}. Hence, Stelle's work encompasses the contributions of quadratic curvature terms. Nevertheless, we call attention to the main differences between \cite{Stelle:1977ry} and the present paper: First, our work is performed in the EC formalism instead of the metric formalism. Theories of gravity in the first order formalism generate two independent field equations, allowing a route to investigate gravitational theories with torsional degrees of freedom. Moreover, the inclusion of fermions as external spin source can also be considered in this formalism. Second, as already explained, our perturbed solution is a deformation around the Schwarzschild-de Sitter spacetime obtained by considering the quadratic curvature term as a perturbation while Stelle's result is a perturbed solution around the Minkowski background obtained by imposing a perturbation on the solution. Similarly, \cite{Nojiri:2013su}, a perturbative solution around the Nariai solution \cite{Nariai1,Nariai2} is found in $f(R)$ gravities in the metric formalism. In \cite{Cognola:2015uva,Cognola:2015wqa}, also in the metric formalism, static solutions of $f(R)\propto R^2$ gravities are studied and the associated thermodynamics are explored in a similar way as ours.

It is worth mentioning that, although our main motivation is the action originated as an emergent gravity in a quantum gauge theory scenario \cite{Sobreiro:2011hb,Sobreiro:2012dp,Sobreiro:2012iv,Sobreiro:2016fks}, our analysis and discussions are general enough to encode any gravity theory as above described. Nevertheless, we will refer to the model \cite{Sobreiro:2011hb,Sobreiro:2012dp,Sobreiro:2010qf,Sobreiro:2016fks} whenever we find it elucidative. It is important to comment about the known problem of unitarity that generally plagues theories of gravity with higher order derivatives. This is discussed already in \cite{Stelle:1977ry}. However, this problem relies at quantum level and is not relevant for effective theories of gravity such as \cite{Sobreiro:2011hb,Sobreiro:2012dp,Sobreiro:2012iv,Sobreiro:2016fks} where the gravity action we consider in the present work is derived as an effective theory from a Yang-Mills theory with $SO(5)$ gauge symmetry. It may also be mentioned that, even in the metric formulation, there are special cases of higher derivative gravities where the problem of non-unitarity can be avoided, see for instance \cite{1987qftq....2..129H,Hawking:2001yt}. On the other hand, there are some interesting results showing that the unitarity problem might not be immediate when connections are treated independently \cite{Alvarez:2017spt,Alvarez:2017ayn,Alvarez:2018lrg}.

Perhaps, the most important general motivation would be the recent detection of gravitational waves by the \textit{Laser Interferometer Gravitational-wave Observatory} (LIGO) collaboration \cite{Abbott:2016blz}. Such discover brings a new era on black hole physics as well as the possibility to test alternative solutions derived of modified theories of gravity that should expand the horizons of the well-known Einstein's gravitational theory. From the point of view of the gravity model in \cite{Sobreiro:2011hb}, this is a motivation for a long term work on the possibility of implementing new tests to such induced gravity. Moreover, the relevance of the results of the present work may also be useful beyond gravity itself and can be applied to analogue models. For instance, the new perturbative solution found here was already employed in the study sound waves patterns in transonic regimes flowing at a de Laval nozzle \cite{daRocha:2017tiz}.

This work is organized as follows: In Sec.~\ref{odesys} we define the model and the respective field equations and specify them for static and spherically symmetric variables. The exact solution is discussed in Sect.~\ref{exacsol}. In Sec.~\ref{pertsol} we provide the perturbative solution. In Sect.~\ref{therm}, a the thermodynamical aspects of the solutions are briefly discussed. Finally, our final considerations are displayed in Sect.~\ref{rmks}.

\section{Action and field equations}\label{odesys}

In \cite{Sobreiro:2011hb} an effective theory of gravity originating from a gauge theory was built. The construction, in summary, considers an initial pure non-Abelian gauge theory with proper degrees of freedom and, under certain energy conditions, this theory ends up in an induced gravity -- more details can be found in \cite{Sobreiro:2011hb,Sobreiro:2012dp}. Consequently, from there we can stand the following gravity action:
\begin{equation}\label{ym-map-grav-obs}
 S_{\mathrm{grav}}=\frac{1}{16\pi G}\int \bigg(\frac{3}{2\Lambda^2}R^{\mathfrak{a}}_{~\mathfrak{b}}\star R_{\mathfrak{a}}^{~\mathfrak{b}} + T^\mathfrak{a}\star T_\mathfrak{a}-\frac{1}{2}\varepsilon_\mathfrak{abcd} R^\mathfrak{ab} e^\mathfrak{c}e^\mathfrak{d} + \frac{\tilde{\Lambda}^2}{12}\varepsilon_\mathfrak{abcd}e^\mathfrak{a}e^\mathfrak{b}e^\mathfrak{c}e^\mathfrak{d}\bigg)~.
\end{equation}
where $R^\mathfrak{a}_{~\mathfrak{b}} = \mathrm{d} \omega^\mathfrak{a}_{~\mathfrak{b}}+\omega^{\mathfrak{a}}_{~\mathfrak{c}} \omega^\mathfrak{c}_{~\mathfrak{b}}$ is the curvature 2-form, $T^\mathfrak{a}=\mathrm{d}e^\mathfrak{a}+\omega^\mathfrak{a}_{~\mathfrak{b}}e^\mathfrak{b}$ is the torsion 2-form, $\omega^\mathfrak{a}_{~\mathfrak{b}}$ is the spin connection 1-form, and $e^\mathfrak{a}$ is the vierbein 1-form. Moreover, $\star$ stands for the Hodge dual operator, $G$ is the Newton's constant, $\widetilde{\Lambda}^2$ is the cosmological constant, and $\Lambda^2$ is a mass parameter. The corresponding vacuum field equations are obtained from the usual variational methods. Varying the action \eqref{ym-map-grav-obs} with respect to the vierbein, then we get the first field equation, 
\begin{equation} \label{eq:eq_mov1}
\frac{3}{2\Lambda^2}R^{\mathfrak{bc}} \star (R_{\mathfrak{bc}}e_{\mathfrak{a}})+ T^{\mathfrak{b}}\star\left(T_{\mathfrak{b}}e_{\mathfrak{a}}\right)+\mathrm{D}\star T_{\mathfrak{a}}- \varepsilon_{\mathfrak{abcd}}\left( R^{\mathfrak{bc}}e^{\mathfrak{d}} - \frac{\tilde{\Lambda}^{2}}{3}e^{\mathfrak{b}}e^{\mathfrak{c}}e^{\mathfrak{d}}\right) = 0~.
\end{equation}
The variation of the action \eqref{ym-map-grav-obs} with respect to the spin connection yields
\begin{equation} \label{eq:eq_mov2}
\frac{3}{\Lambda^2}D\star R_{\mathfrak{a}\mathfrak{b}} 
+e_{\mathfrak{b}} \star T_{\mathfrak{a}}
-e_{\mathfrak{a}} \star T_{\mathfrak{b}} -\varepsilon_{\mathfrak{abcd}} T^{\mathfrak{c}} e^{\mathfrak{d}} = 0~,
\end{equation}
where $D=\mathrm{d}+\omega$, the exterior covariant derivative. Eqs.~\eqref{eq:eq_mov1} and \eqref{eq:eq_mov2} are coupled nonlinear differential equations and thus, highly difficult to solve without any special insight. For this reason, we proceed with the simplest case where torsion is set to zero. Since we are not considering spin sources, this simplification is not harmful, but sufficiently straight in order to predict extended aspects in the results when compared to Einstein’s gravity. Hence, Eqs.~\eqref{eq:eq_mov1} and \eqref{eq:eq_mov2} reduce to
\begin{eqnarray}
\frac{3}{2\Lambda^2}R^{\mathfrak{bc}} \star (R_{\mathfrak{bc}}e_{\mathfrak{a}}) - \varepsilon_{\mathfrak{abcd}}\left(R^{\mathfrak{bc}} e^{\mathfrak{d}} - \frac{\tilde{\Lambda}^{2}}{3}e^{\mathfrak{b}}e^{\mathfrak{c}}e^{\mathfrak{d}}\right) &=& 0~, \label{eq:eq_mov1mod}\\
\frac{3}{\Lambda^2}D\star R_{\mathfrak{a}\mathfrak{b}} &=& 0~.\label{eq:eq_mov2mod}
\end{eqnarray}
It is worth mentioning that the action \eqref{ym-map-grav-obs} describes an emergent gravity associated to an $SO(5)$ Yang-Mills theory \cite{Sobreiro:2011hb}. However, the field equations \eqref{eq:eq_mov1mod} and \eqref{eq:eq_mov2mod} are general enough to describe any gravity theory with a Riemann squared curvature term in the action. Whenever relevant, we will comment about the model developed in \cite{Sobreiro:2011hb} and the results here found. It should be clear that, in \eqref{eq:eq_mov1mod}, the last two terms correspond to the usual Einstein equation with cosmological constant. On the other hand, \eqref{eq:eq_mov2mod} is not present in the usual Einstein-Hilbert equations since it comes from the quadratic curvature term of the action.

Our aim is to solve equations  \eqref{eq:eq_mov1mod} and  \eqref{eq:eq_mov2mod} for static and spherically symmetric conditions. To do so, we will consider two different situations:
\begin{itemize}
\item First, we consider both Eqs.~\eqref{eq:eq_mov1mod} and  \eqref{eq:eq_mov2mod} and find an exact solution, corresponding to a strong influence of the quadratic curvature term\footnote{In the exact solution, the quadratic curvature term is assumed to be strong rather than a small perturbation around the Einstein-Hilbert term, \emph{i.e.}, this term is comparable to the Einstein-Hilbert term.}. The result we find is the usual de Sitter solution with an effective cosmological constant given by a mix between $\Lambda^2$ and $\widetilde{\Lambda}^2$.

\item Second, by taking $\Lambda^2$ as a huge quantity\footnote{This situation is consistent with the results of \cite{Sobreiro:2016fks,Assimos:2013eua} where 1- and 2-loop explicit computations predict a huge value for $\Lambda$ in the effective gravity model constructed in \cite{Sobreiro:2011hb}.} when compared to the quadratic curvature term, we treat this term as a perturbation. The perturbed solution is a deformation of the usual Schwarzschild-de Sitter solution. To obtain such solution, the equation \eqref{eq:eq_mov2mod} is neglected because it is a pure perturbation in $\Lambda^{-2}$. This situation is equivalent to set $T=0$ at the action \eqref{ym-map-grav-obs} before the computation of the field equations.
\end{itemize}

Eqs.~\eqref{eq:eq_mov1mod} and \eqref{eq:eq_mov2mod}, in Schwarzschild coordinates,
\begin{equation}\label{spheric-symm}
e^0=e^{\alpha(r)}\mathrm{d}t~~~,~~~e^1=e^{\beta(r)}\mathrm{d}r ~~~,~~~e^2=r\mathrm{d}\theta~~~,~~~e^3=r \sin\theta \mathrm{d}\phi~,
\end{equation}
can be recasted as
\begin{equation}\label{edo-t}
\sigma\left[2\left(\frac{e^{-2\beta}\dot{\beta}}{r}\right)^2+\left(\frac{1-e^{-2\beta}}{r^2}\right)^2\right]+2\left(\frac{e^{-2\beta}\dot{\beta}}{r}\right)+\frac{1-e^{-2\beta}}{r^2}+3\lambda =0~,
\end{equation}
\begin{equation}\label{edo-r}
\sigma\left[2\left(\frac{e^{-2\beta}\dot{\alpha}}{r}\right)^2+\left(\frac{1-e^{-2\beta}}{r^2}\right)^2\right]-2\left(\frac{e^{-2\beta}\dot{\alpha}}{r}\right)+\frac{1-e^{-2\beta}}{r^2}+3\lambda =0~,
\end{equation}
\begin{eqnarray}\label{edo-theta-phi}
& &\sigma e^{-4\beta}\left[(\ddot{\alpha}+\dot{\alpha}^2-\dot{\alpha}\dot{\beta})^2+\frac{1}{r}\left(\dot{\alpha}^2+\dot{\beta}^2\right)\right] -e^{-2\beta}\left[\ddot{\alpha}+\dot{\alpha}^2-\dot{\alpha}\dot{\beta}+\frac{1}{r}\left(\dot{\alpha}-\dot{\beta}\right)\right]+3\lambda=0~,
\end{eqnarray}
\begin{equation}\label{edo-scalarcurv}
\dot{\mathcal{R}}=0~,
\end{equation}
for $\mathfrak{a}=0~$, $\mathfrak{a}=1~$ and $\mathfrak{a}=2$, respectively. In Eq.~\eqref{edo-scalarcurv}, $\mathcal{R}$ stands for the scalar curvature. The dot notation indicates derivatives respect to $r$, since $\alpha\equiv\alpha(r)$, $\beta\equiv\beta(r)$ and $\mathcal{R}$ depend only on this variable. We notice that differential equations obtained for $\mathfrak{a}=2$ and $\mathfrak{a}=3$ are identical. The constants in Eqs.~\eqref{edo-t}, \eqref{edo-r} and \eqref{edo-theta-phi} are $\sigma\equiv -3/(2\Lambda^2)$ and $\lambda\equiv -\tilde{\Lambda}^2/3$.
The combination of Eqs.~\eqref{edo-t} and \eqref{edo-r} leads to the following constraint
\begin{equation}
\left(\dot{\alpha}+\dot{\beta}\right) \left(\dot{\alpha}-\dot{\beta} +\frac{r}{\sigma}e^{2\beta}\right)=0~,
\end{equation}
which allows two possibilities:
\begin{itemize}
\item $\dot{\alpha}+\dot{\beta}=0 \Rightarrow \alpha+\beta=f(t)$ 

Thus, we have freedom to re-scale the time coordinate with simply $f(t)=0$.

\item $\dot{\alpha}-\dot{\beta} +\sigma^{-1}r\exp(2\beta)=0$

This is a coupled nonlinear differential equation for $\alpha$ and $\beta$ which is not the usual condition appearing in the Literature.
\end{itemize}
We will stick to the first possibility, which is the usual relation in General Relativity and impose $\alpha=-\beta$. Of course, the second possibility could bring interesting aspects and will, certainly, be investigated in a future work.

\section{Exact solution}\label{exacsol}

In order to solve analytically the system of equations \eqref{edo-t}-\eqref{edo-scalarcurv}, we start by subtracting Eq.~\eqref{edo-r} from Eq.~\eqref{edo-theta-phi}, obtaining\footnote{The partial derivatives were changed to ordinary ones, since $\beta$ is only $r$-dependent.},
\begin{equation}\label{nde-tminustheta}
\left(\frac{\ddot{h}}{2}+\frac{1-h}{r^2}\right)\left[\sigma\left(\frac{\ddot{h}}{2}-\frac{1-h}{r^2}\right)-1\right]=0~,
\end{equation}
where, the condition $\alpha+\beta=0$ was employed. Moreover, we have defined $e^{-2\beta}\equiv h$ and $\dot{h}\equiv dh/dr$. Eq.~\eqref{nde-tminustheta} can be decomposed in two independent differential equations where only one must be zero. First, we have
\begin{equation}\label{exact-edo01}
r^2\ddot{h}+2h-2\left(1+\frac{r^2}{\sigma}\right)=0~,
\end{equation}
whose solution is
\begin{equation}
h(r)=1+\frac{r^2}{2\sigma}+\sqrt{r}\left[c_1\cos\left(\frac{\sqrt{7}}{2}\ln r\right)+c_2\sin\left(\frac{\sqrt{7}}{2}\ln r\right)\right]~,
\end{equation}
where $c_1$ and $c_2$ are integration constants. It turns out that this solution does not satisfy the whole differential equation system \eqref{edo-t}-\eqref{edo-scalarcurv}.

The second possibility is
\begin{equation}\label{exact-edo02}
r^2\ddot{h}-2h+2=0~,
\end{equation}
whose solution is given by
\begin{equation}
h(r)=1+c_3r^2+\frac{c_4}{r}~,\label{hhhh}
\end{equation}
where $c_3$ and $c_4$ are integration constants. It is a straightforward computation to show that we must have $c_4=0$ and $c_3\neq0$ in order to the solution \eqref{hhhh} satisfy the system \eqref{edo-t}-\eqref{edo-scalarcurv}. Moreover, there are two possible values for the constant $c_3$, namely $\Upsilon_p$ and $\Upsilon_m$,
\begin{eqnarray}\label{Ypm}
\Upsilon_p &=& \frac{\Lambda^2}{3} \left[1 + \sqrt{1-2\frac{\tilde{\Lambda}^2}{\Lambda^2}}\right]~,\nonumber\\
\Upsilon_m &=& \frac{\Lambda^2}{3} \left[1 - \sqrt{1-2\frac{\tilde{\Lambda}^2}{\Lambda^2}}\right]~.
\end{eqnarray}
Hence
\begin{eqnarray}\label{exact-sol-pm}
e^{-2\beta_p} &=& 1-\Upsilon_p r^2~,\nonumber\\
e^{-2\beta_m} &=& 1-\Upsilon_m r^2~.
\end{eqnarray}
The solutions \eqref{exact-sol-pm} satisfy the system of differential equations \eqref{edo-t}-\eqref{edo-scalarcurv}, simultaneously. This is an important and necessary verification since this system of equations is over-determined. From \eqref{Ypm}, it is clear that $\tilde{\Lambda}^2$ cannot exceed $\Lambda^2/2$, otherwise, the solution founded is inconsistent. Moreover, if $2\tilde{\Lambda}^2=\Lambda^2$, only one solution is allowed $\Upsilon_m=\Upsilon_p$.

Now, if we assume that $\Lambda^2$ has a large value and $\tilde{\Lambda}^2$ has a small value\footnote{This is consistent, for instance, with the explicit values $\Lambda^2\approx 7.665\times 10^{31}\textrm{TeV}^2\gg\tilde{\Lambda}^2\approx 1.000\times 10^{-92}\textrm{TeV}^2$ found in \cite{Assimos:2013eua,Sobreiro:2016fks}.}, we can expand \eqref{Ypm} to find
\begin{eqnarray}\label{Yappr}
\Upsilon_s &\approx &\frac{1}{3}\tilde{\Lambda}^2~,\nonumber\\
\Upsilon_b &\approx &\frac{2}{3}\Lambda^2~.
\end{eqnarray}
The first case, $\Upsilon_s$ has a narrow value if we take $\widetilde{\Lambda}$ as its observational value. On the other hand, the second case stems for a de Sitter-like space with a very small radius, since $\Lambda^2$ is very big. Hence, we have a weak curvature regime for $\Upsilon_s$ and a strong one for $\Upsilon_b$.

Obviously, all usual properties of de Sitter spacetime are maintained\footnote{An alternative and simple way to find the solution \eqref{exact-sol-pm} is to directly deal with equation \eqref{eq:eq_mov1} in form notation and try an \emph{ansatz} solution of the form
\begin{equation}
R^{\mathfrak{ab}}=\zeta e^{\mathfrak{a}} e^{\mathfrak{b}}\;,\label{ap1}
\end{equation}
where $\zeta$ is a constant mass parameter. The solution \eqref{ap1} is a natural choice since we have the usual cosmological constant term in \eqref{eq:eq_mov1}. The direct substitution of \eqref{ap1} in \eqref{eq:eq_mov1} leads to the characteristic equation for $\zeta$,
\begin{equation}
\frac{3}{2\Lambda^2}\zeta^2-\zeta+\frac{\tilde{\Lambda}^2}{3}=0\;,\label{ap2}
\end{equation}
providing $\zeta=\Upsilon_{p,m}$. Hence, solution \eqref{ap1} is an alternative covariant form of the curvature associated with solution \eqref{exact-sol-pm}.}.

\section{Perturbative solution}\label{pertsol}

From this point, we consider the quadratic curvature to be a small perturbation in Eq.~\eqref{edo-t}. For that, we multiply Eq.~\eqref{edo-t} by $\lambda$, then
\begin{equation}\label{edo-t-pert}
\eta\left[2\left(\frac{e^{-2\beta}\dot{\beta}}{r}\right)^2+\left(\frac{1-e^{-2\beta}}{r^2}\right)^2\right]+\lambda\left[2\left(\frac{e^{-2\beta}\dot{\beta}}{r}\right)+\frac{1-e^{-2\beta}}{r^2}+3\lambda\right] =0~.
\end{equation}
In this form, Eq.~\eqref{edo-t-pert} can be solved analytically through the employment of perturbation theory if $\eta\equiv\sigma\lambda\equiv\tilde{\Lambda}^2/2\Lambda^2$ is a very small dimensionless parameter. Hence, the quadratic term can be treated as a perturbation around the term proportional to $\lambda$. It is evident that the term proportional to $\lambda$ is the usual Einstein equation with cosmological constant $\widetilde{\Lambda}^2$. For simplicity, let $u(r)=1-e^{-2\beta}$ and take all derivatives as ordinary ones. So, we rewrite Eq.~\eqref{edo-t-pert} as
\begin{equation}\label{edo-t-u}
\eta\left[\frac{1}{2}{\dot{u}}^2+\left(\frac{u}{r}\right)^2\right] + \lambda\left(r\dot{u} + u + 3\lambda r^2\right)=0~.
\end{equation}
A perturbative solution of Eq.~\eqref{edo-t-u} has the general form
\begin{equation}\label{SPert-geral}
u(r)=u_0(r) + \eta u_1(r) + \eta^2 u_2(r) + \eta^3 u_3(r) + \cdots~.
\end{equation}
Substituting \eqref{SPert-geral} in the Eq.~\eqref{edo-t-u}, and splitting order by order in $\eta$, we find an infinite set of iterated differential coupled equations.
\begin{eqnarray}\label{hierarq-pert-u}
\dot{(r u_0)}+3\lambda r^2 &=& 0~,\nonumber\\
\dot{(r u_1)} + \frac{1}{2\lambda}\dot{u}_0^2 + \frac{u_0^2}{\lambda r^2} &=& 0~,\nonumber\\
\dot{(r u_2)}+\frac{1}{\lambda}\left(\dot{u}_0 \dot{u}_1\right) + \frac{2u_0 u_1}{\lambda r^2} &=& 0~,\nonumber\\
\dot{(r u_3)} + \frac{1}{2\lambda}\dot{u}_1^2 + \frac{u_1^2}{\lambda r^2} +\frac{1}{\lambda}\left(\dot{u}_0\dot{u}_2\right) + \frac{2u_0 u_2}{\lambda r^2} &=& 0~,\nonumber\\
\dot{(r u_4)} + \frac{1}{\lambda}\left(\dot{u}_0\dot{u}_3\right) + \frac{2u_0 u_3}{\lambda r^2} + \frac{1}{\lambda}\left(\dot{u}_1\dot{u}_2\right) + \frac{2u_1 u_2}{\lambda r^2} &=& 0~,\nonumber\\
&\vdots&\;.
\end{eqnarray}
With such hierarchy of equations \eqref{hierarq-pert-u}, we can solve, iteratively, all the equations above. Starting with the zeroth order, we obtain
\begin{equation}
u_0 = \frac{\tilde{\Lambda}^2}{3} r^2 + \frac{2GM}{r}~,
\end{equation}
which is the usual Schwarzschild-de Sitter solution \cite{Gibbons:1977mu,Cardoso:2003sw,Faraoni:2015ula}, obviously, since this is the situation for $\eta=0$. Hence, the integration constant at the $1/r$ term is obtained from the Newtonian limit -- see Appendix~\ref{NewtonLIM} for a detailed discussion. Solving iteratively the rest of the equations \eqref{hierarq-pert-u}, we find the following solution (explicitly only at fourth order), 
\begin{eqnarray}\label{pert-sol-4th}
e^{-2\beta}&=& 1-\frac{2GM}{r}-\frac{\tilde{\Lambda}^2}{3}r^2-\eta\left(\frac{\mathcal{C}_{12}}{r}+\mathcal{C}_{11}r^2+\frac{\mathcal{C}_{13}}{r^4}\right)- \eta^2\left(\frac{\mathcal{C}_{22}}{r}+\mathcal{C}_{21}r^2+\frac{\mathcal{C}_{23}}{r^4}+\frac{\mathcal{C}_{24}}{r^7}\right)\nonumber\\
&-&\eta^3\left(\frac{\mathcal{C}_{32}}{r}+\mathcal{C}_{31}r^2+\frac{\mathcal{C}_{32}}{r^4}+\frac{\mathcal{C}_{34}}{r^7}+\frac{\mathcal{C}_{35}}{r^{10}}\right) - \eta^4\left(\frac{\mathcal{C}_{42}}{r}+\mathcal{C}_{41}r^2+\frac{\mathcal{C}_{43}}{r^4}+\frac{\mathcal{C}_{44}}{r^7}+\frac{\mathcal{C}_{45}}{r^{10}}+\frac{\mathcal{C}_{46}}{r^{13}}\right)+\ldots\nonumber\\
\end{eqnarray}
where the constants $\mathcal{C}_{k\ell}$ can be arranged as
\begin{equation}
\mathcal{C}_{k\ell}\equiv\left(\begin{array}{cccccc}\label{CIs}
\frac{\tilde{\Lambda}^2}{3} & 2GM  &  & & & \\
\frac{\tilde{\Lambda}^2}{3} & \mathcal{C}_{12}  & \frac{6G^2M^2}{\tilde{\Lambda}^2} & & & \\
2\frac{\tilde{\Lambda}^2}{3} & \mathcal{C}_{22} & \frac{6GM}{\tilde{\Lambda}^2}\Omega_1 & -\frac{36G^3M^3}{\tilde{\Lambda}^4} &  & \\
5\frac{\tilde{\Lambda}^2}{3} & \mathcal{C}_{32} & \frac{9}{2\tilde{\Lambda}^2}\Omega_2 & \frac{54G^2M^2}{\tilde{\Lambda}^4}\Omega_4 & \frac{312G^4M^4}{\tilde{\Lambda}^6} & \\
14\frac{\tilde{\Lambda}^2}{3} & \mathcal{C}_{42}  &  -\frac{3}{\tilde{\Lambda}^2}\Omega_3 & \frac{3GM}{\tilde{\Lambda}^4}\Omega_5 & \frac{54G^2M^2}{\tilde{\Lambda}^6}\Omega_6 & -\frac{3564G^5M^5}{\tilde{\Lambda}^8}
\end{array}\right)
\end{equation}
where the index $k$ (line) and $\ell$ (column) run along the discrete intervals $[0,\infty]$ and $[1,k+2]$, respectively, and
\begin{eqnarray}
\Omega_1 &=& \mathcal{C}_{12}-2GM~,\nonumber\\
\Omega_2 &=& \left[\mathcal{C}_{12}^2 + 4GM\left(6GM-2\mathcal{C}_{12}-\mathcal{C}_{22}\right)\right]~,\nonumber\\
\Omega_3 &=&\left[\mathcal{C}_{12}\left(\mathcal{C}_{12}+ \mathcal{C}_{22}-12GM\right)- 2GM\left(2\mathcal{C}_{22}+ \mathcal{C}_{32}-20GM\right)\right]~,\nonumber\\
\Omega_4 &=& \left(8GM-3\mathcal{C}_{12}\right)~,\nonumber\\
\Omega_5 &=& \left[3\mathcal{C}_{12}^2-2GM\left( 12\mathcal{C}_{12}+3\mathcal{C}_{22}-24GM\right)\right]~,\nonumber\\
\Omega_6 &=& 2\left(\mathcal{C}_{12}-3GM\right)~.
\end{eqnarray}
In fact, the solution \eqref{pert-sol-4th} can be generalized to all orders in a very concise form given by
\begin{equation}\label{pert-sol-gen}
e^{-2\beta}=1-\sum_{k=0}^{\infty}\eta^k\sum_{\ell=1}^{k+2}
\mathcal{C}_{k\ell}r^{5-3\ell}~~,
\end{equation}
where the general constants $\mathcal{C}_{k\ell}$ depend on the previous order constants. In particular, the $\mathcal{C}_{k2}$, with $k=0,1,2,\dots$, stand for the actual integration constants.

A first comment about the perturbative solution Eq.~\eqref{pert-sol-4th} is that it is possible to redefine the mass as an effective one in Eq.~\eqref{pert-sol-4th} by collecting all coefficients proportional to $1/r$. Hence, an effective Schwarzschild term arises as $2G\tilde{M}/r$, where $2G\tilde{M}=2GM+\eta \mathcal{C}_{12}+ \eta^2 \mathcal{C}_{22} + \eta^3 \mathcal{C}_{32}+\cdots$.

A second important issue to be considered is the convergence of the perturbative solution Eq.~\eqref{pert-sol-4th} in the form Eq.~\eqref{pert-sol-gen}. The convergence of the series is not a straightforward issue since the solution \eqref{pert-sol-4th} carries arbitrary integration constants $\mathcal{C}_{k2}$ in a sequence of functions related to a single variable $r$ and the expansion parameter $\eta$. The expansion parameter is assumed to be small, but with a certain freedom to fix it, in principle, inside the interval $(0,1)$. Nevertheless, establishing constraints on the values of the constants $\mathcal{C}_{k\ell}$ aggregated with the domain of validity of $r$ and the small value of the parameter of the expansion $\eta$ we endorse that the series is uniformly convergent -- See Appendix~\ref{CSdemo} for the details. The conditions encountered are summarized by
\begin{eqnarray}
\eta<\frac{1}{4}\;,\nonumber\\
\left|\frac{\mathcal{C}_{(k+1)\ell}}{\mathcal{C}_{k\ell}}\right|&<&\frac{1}{\eta}\;,\;\;\;\;\forall\; k\in[0,\infty)\;\;\; \mathrm{and}\;\;\; \ell\in[2,\infty)\nonumber\\
\frac{\left|\sum_{k=\ell-1}^\infty \eta^k\mathcal{C}_{k(\ell+1)}\right|}{\left|\sum_{k=\ell-2}^{\infty} \eta^k\mathcal{C}_{k\ell}\right|}&<&1\;, \;\;\;\;\forall\; k\in[0,\infty)\;\;\; \mathrm{and}\;\;\; \ell\in[3,\infty).\label{convergence00}
\end{eqnarray}
The first condition in \eqref{convergence00} establishes that the naive interval $(0,1)$ for $\eta$ is reduced to the smaller interval $(0,1/4)$. The second condition in \eqref{convergence00} says that the smaller is $\eta$, the bigger is the acceptable ratio between two $k$-consecutive constants $\mathcal{C}_{k\ell}$. Combining this second condition with the first one, we find that the critical situation ($\eta=1/4$) is $\left|\frac{\mathcal{C}_{(k+1)\ell}}{\mathcal{C}_{k\ell}}\right|<4$. The third condition in \eqref{convergence00} is more elaborated and provides an extra constraint amongst all constants for $\ell\in[3,\infty]$.

A third point refers to the differential equation system, \eqref{edo-t}--\eqref{edo-theta-phi} with the assumption $\alpha=-\beta$ arising from the combination of Eqs.~\eqref{edo-t} and \eqref{edo-r}. The solution \eqref{pert-sol-gen} is, formally, the solution of the system \eqref{edo-t}--\eqref{edo-r}. However, we can mention that Eq.~\eqref{edo-theta-phi} is also satisfied. In fact, one can employ the same perturbative technique to solve Eq.~\eqref{edo-theta-phi} with the condition $\alpha=-\beta$ and obtain a similar result to Eq.~\eqref{pert-sol-gen}, up to a redefinition of the integration constants -- See Appendix~\ref{EDO9} for more details.

Simple consistency checks of the solution \eqref{pert-sol-gen} are: the limits $\eta\rightarrow 0$ and $\tilde{\Lambda}^2=0$, providing a pure Schwarzschild solution; the limits $M=0$ and $\eta\rightarrow 0$ result in a de Sitter spacetime solution; and, as already seen, the limit $\eta\rightarrow 0$ provides the Schwarzschild-de Sitter solution.

It is interesting to take the limit $r\gg 2GM$ in the solution \eqref{pert-sol-4th}. The result is immediate,
\begin{equation}\label{pert-ds-far}
e^{-2\beta}\approx\left(1-\tilde{\Upsilon} r^2\right)~,
\end{equation}
where
\begin{equation}\label{Y-far}
\tilde{\Upsilon}\approx \frac{\tilde{\Lambda}^2}{3}+\eta\left(\frac{\tilde{\Lambda}^2}{3}\right)+\eta^2\left(2\frac{\tilde{\Lambda}^2}{3}\right)+\eta^3\left(5\frac{\tilde{\Lambda}^2}{3}\right)+\eta^4\left(14\frac{\tilde{\Lambda}^2}{3}\right)+\dots~,
\end{equation}
which is a perturbatively asymptotically de Sitter spacetime, as expected. Remarkably, the relation \eqref{Y-far} can be written as
\begin{equation}\label{Y-notrunc}
\widetilde{\Upsilon}=\left(\sum_{w=0}^\infty \eta^w c_w\right)\frac{\tilde{\Lambda}^2}{3}~,
\end{equation}
where
\begin{equation}\label{catalanNum}
c_w=\frac{(2w)!}{(w+1)!w!}
\end{equation}
are the Catalan numbers\footnote{Named after the discovery of the sequence of natural numbers by the Belgian mathematician Eug\`ene C. Catalan $(1814-1894)$, which made several contributions to combinatorial mathematics \cite{Weisstein}.}. Moreover, by setting $\eta=0$ in \eqref{Y-far} we find $\tilde{\Upsilon}=\Upsilon_s$. Hence, we have an asymptotic relation between the perturbative solution and the exact one.  Further, by expanding the constant $\Upsilon_p$ in \eqref{Ypm} for small $\widetilde{\Lambda}^2/\Lambda^2$, we get the same expression \eqref{Y-notrunc}. Hence, we have consistency between the exact and perturbative solution for small $\widetilde{\Lambda}^2/\Lambda^2$.

For the next sections, for the sake of simplicity, we keep up to the first order in $\eta$. Thus, the solution \eqref{pert-sol-4th} is truncated to 
\begin{equation}\label{pert-sol-1st} 
e^{-2\beta} \approx 1-\frac{\mathcal{C}_{02}}{r}-\mathcal{C}_{01}r^2-\eta\left(\frac{\mathcal{C}_{12}}{r}+\mathcal{C}_{11}r^2+\frac{\mathcal{C}_{13}}{r^4}\right)~.
\end{equation}
where the explicit form of the constants are displayed in \eqref{CIs}.

\subsection{Horizons}\label{ehor1}

Eq.~\eqref{pert-sol-1st} can be used to determine the horizons by solving $e^{-2\beta}=0$
\cite{Cardoso:2003sw,Faraoni:2015ula,Bhattacharya:2010vr,Wald:1984rg}. Thus, we must solve the following perturbed algebraic equation
\begin{equation}\label{eq-hor-1st}
r^3\left(r-\mathcal{C}_{02}-\mathcal{C}_{01}r^3\right)
-\eta\left(\mathcal{C}_{11}r^6+\mathcal{C}_{12}r^3 + \mathcal{C}_{13}\right)=0~,
\end{equation}
where the constants $\mathcal{C}$'s are listed in \eqref{CIs}. The solution can be taken as a perturbative one of the form
\begin{equation}\label{r-hor-1st}
r\approx r_0+\eta r_1~.
\end{equation}
Substituting Eq.~\eqref{r-hor-1st} in Eq.~\eqref{eq-hor-1st} we obtain a system of two algebraic equations. At zeroth order, already with the substitution of constants $\mathcal{C}$s, we have
\begin{equation}\label{cubeq-0}
r_0^3-\frac{3}{\tilde{\Lambda}^2}r_0+\frac{6GM}{\tilde{\Lambda}^2}=0~,
\end{equation}
and, at first order,
\begin{equation}\label{cubeq-1}
r_0^3\left(1+3\mathcal{C}_{01}r_0^2\right)r_1-\mathcal{C}_{13}r_0^3-\mathcal{C}_{11}r_0^6 - \mathcal{C}_{13}=0~,
\end{equation}
which we left with the constants $\mathcal{C}$'s for the sake of simplicity. The polynomial discriminant of the Eq.~\eqref{cubeq-0} is easily computed
\begin{equation}
\Delta=\frac{108}{\tilde{\Lambda}^6}\left(1-9G^2M^2\tilde{\Lambda}^2\right)~,
\end{equation}
which is important to determine the nature of the roots of Eq.~\eqref{cubeq-0}. If, and only if $\Delta>0$, Eq.~\eqref{cubeq-0} has three real roots. Such condition implies that $3GM\tilde{\Lambda}<1$, since $\tilde{\Lambda}$, $G$ and $M$ are positive quantities. By applying the trigonometric method to find all the roots of Eq.~\eqref{cubeq-0}, it is found only two different positive roots, namely,
\begin{eqnarray}\label{Hroots-0}
r_{01} &=& \frac{1}{\tilde{\Lambda}}\left(c_\xi+\sqrt{3}s_\xi\right)~,\nonumber\\
r_{02} &=& \frac{1}{\tilde{\Lambda}}\left(c_\xi-\sqrt{3}s_\xi\right)~,
\end{eqnarray}
where $c_\xi\equiv\cos\xi$, $s_\xi\equiv\sin\xi$ and $\xi=1/3\arccos\left(3GM\tilde{\Lambda}\right)$. The third root is $r_{03}=-\left(r_{01}+r_{02}\right)$, which is essentially negative and it is not physical. Since $0<3GM\tilde{\Lambda}<1\;\Rightarrow\;0<\arccos\left(3GM\tilde{\Lambda}\right)<\pi/2$ we have $r_{01}>r_{02}>0$. Accordingly, $r_{01}$ stands for the cosmological horizon and $r_{02}$ stands for the mass distribution event horizon. Further, when $s_\xi=0$, which correspond to $M=0$ or $3GM\tilde{\Lambda}=1$, $n\in\{0,1,2,\dots\}$, we have two coincident horizons, $i.e.$, $r_b=r_c$.

\begin{figure}[htb]
	\centering
		\includegraphics[width=0.6\textwidth]{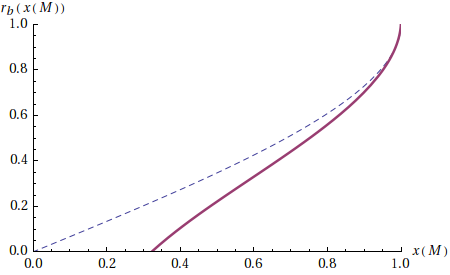}
		\caption{ Event horizon related to the mass spherical distribution.
		$r_b(x(M))$ is in units of $\tilde{\Lambda}^{-1}$ and $3GM\tilde{\Lambda}\equiv x $.
		The dashed curve represents the event horizon behaviour of a standard black hole and the thick curve represents the perturbative horizon.
		For the thick curve we adopted $\eta=10^{-1}$.}
\label{graphrb}
\end{figure}

\begin{figure}[htb]
  \centering
   \includegraphics[width=0.6\textwidth]{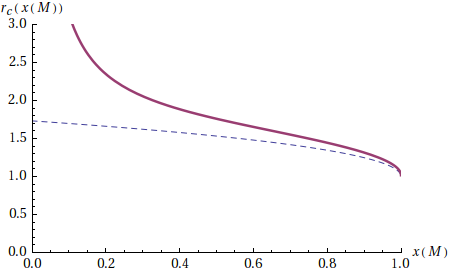}
	\caption{Cosmological horizon.
			$r_c(x(M))$ is in units of $\tilde{\Lambda}^{-1}$, $3GM\tilde{\Lambda}\equiv x$ and $\eta=10^{-1}$.
			The dashed curve represents the cosmological horizon obtained from the Schwarzschild-de Sitter geometry.
			The thick curve is the cosmological horizon with the correction from the quadratic term contribution.}
\label{graph-rc}
\end{figure}

Now, the substitution of Eq.~\eqref{Hroots-0} in Eq.~\eqref{cubeq-1} provides
\begin{equation}\label{Hroots-1}
r_{1\ell}=\frac{1}{3\left(1-\frac{\tilde{\Lambda}^2}{3} r_{0\ell}^2\right)} \left(-\frac{6G^2M^2}{\tilde{\Lambda}^2}\frac{1}{r_{0\ell}^3}+\mathcal{C}_{12}+\frac{\tilde{\Lambda}^2}{3} r_{0\ell}^3\right)~,
\end{equation}
with $\ell=1$ or $\ell=2$. Hence, the form of the horizons at first order are\footnote{We point out that the horizons depend on the integration constant $C_{12}$, which may depend explicitly on the mass $M$. The qualitative/quantitative behaviour of such horizons and the respective thermodynamical quantities is directly bounded by $C_{12}$. In the next steps we are assuming that $C_{12}$ has a linear dependence on $M$, so all graphs in this paper are plotted under this assumption. This is a natural hypothesis since $C_{12}$ appear as a factor of the pure Schwarzschild term correction $\sim 1/r$. Nevertheless, it may be ajusted by employing more sophisticated boundary conditions. A complete analysis of this fine tuning is left for future investigation \cite{Silveira:2017xxx}.}
\begin{eqnarray}\label{HZs}
r_b &=& \frac{1}{\tilde{\Lambda}}\left\{\left(c_\xi -\sqrt{3}s_\xi\right)+\eta \frac{\sec(3\xi)}{6}\left[-\frac{18G^2M^2\tilde{\Lambda}^2}{\left(c_\xi-\sqrt{3}s_\xi\right)^2}+3\mathcal{C}_{12} \left(c_\xi-\sqrt{3}s_\xi\right) +  \left(c_\xi-\sqrt{3}s_\xi\right)^4\right]\right\} ~,\nonumber\\
r_c &=& \frac{1}{\tilde{\Lambda}}\left\{\left(c_\xi + \sqrt{3}s_\xi\right)+ \eta \frac{\sec(3\xi)}{6}\left[-\frac{18G^2M^2\tilde{\Lambda}^2}{\left(c_\xi+\sqrt{3}s_\xi\right)^2}+3\mathcal{C}_{12} \left(c_\xi+\sqrt{3}s_\xi\right) +  \left(c_\xi+\sqrt{3}s_\xi\right)^4\right]\right\}~,\nonumber\\
\end{eqnarray}
where $r_b$ is the event horizon and $r_c$ is the cosmological one. The limit $\eta\rightarrow 0$ clearly recovers the two horizons for a Schwarzschild-de Sitter spacetime. The behaviour of $r_b$ and $r_c$ are qualitatively displayed\footnote{The $\eta$ parameter, according to \cite{Assimos:2013eua,Sobreiro:2016fks} is very small, providing no actual difference in the plots. For this reason we overestimate $\eta$ for the sake of comparison. The same observation holds for all the plots in this work.} in Figure~\ref{graphrb} and Figure~\ref{graph-rc}, respectively. We observe that the increasing of $M$ implies in the increasing of $r_b$, as expected. On the other hand, $r_c$ decreases as $M$ increases.

There is a subtle point here about apparent horizons since we are dealing with perturbative solutions, $i.e.$, commonly called perturbed black holes. Indeed, such kind of black holes are explored within a gravitational collapse context. In the case of static vacuum solutions, the apparent and event horizons are coincident. Hence, we opted here simply to call them by event horizons \cite{Frolov:1998wf,Hawking:1973uf,Bronnikov:2012wsj}.

\subsection{Singularities}

It is a straightforward calculation to find the possible singularities in the perturbative solution \eqref{pert-sol-1st} by computing the Kretschmann invariant, the scalar curvature and the Ricci tensor squared, namely,
\begin{eqnarray}\label{K-invar}
\mathcal{R}^{\alpha\beta\gamma\delta} \mathcal{R}_{\alpha\beta\gamma\delta}&=&
\frac{48G^2M^2}{r^6}+\frac{8\tilde{\Lambda}^4}{3}+
\eta\left[\left(C_{12}+GM\right)\frac{48GM}{r^6}+\frac{1440G^3M^3}{\tilde{\Lambda}^2r^9}+ \frac{16\tilde{\Lambda}^2}{3}\right]~,\nonumber\\
\mathcal{R}&=&4\tilde{\Lambda}^2+\eta\left(4\tilde{\Lambda}^2 + \frac{36G^2M^2}{\tilde{\Lambda}^2r^6}\right)\;,\nonumber\\
\mathcal{R}^{\alpha\beta}\mathcal{R}_{\alpha\beta}&=& 4\tilde{\Lambda}^4+\eta\left(\frac{72G^2M^2}{r^6}+8\tilde{\Lambda}^4\right)\;.
\end{eqnarray}
It is then clear that a physical singularity exists at $r=0$, as expected. Interestingly, the perturbative contributions to $\mathcal{R}$ and $\mathcal{R}^{\alpha\beta}\mathcal{R}_{\alpha\beta}$ are singular, while their zeroth order terms are not,
\begin{eqnarray}\label{scalar-ricci-invar}
\lim_{r\rightarrow 0}\mathcal{R}&\rightarrow &\eta\infty\nonumber\\
\lim_{r\rightarrow 0}\mathcal{R}^{\alpha\beta}\mathcal{R}_{\alpha\beta}&\rightarrow &\eta\infty ~.
\end{eqnarray}
The physical singularity expressed by \eqref{K-invar} and \eqref{scalar-ricci-invar} are in agreement with the standard results obtained in the Einsteinian gravity.

\section{Some thermodynamical aspects}\label{therm}

In order to improve our analysis for the solutions found so far, we will express the main thermodynamical quantities related to the horizons. For this goal, we need to compute the surface gravities defined through\footnote{Since the Killing field is just $r$-dependent, we can use this simplified expressions rather than the general formula $\zeta^\mathfrak{a}\nabla_\mathfrak{a}\zeta^\mathfrak{b}= \kappa\zeta^\mathfrak{b}$, where $\zeta^\mathfrak{a}$ is the Killing vector field normal to the horizon.} \cite{Cardoso:2003sw}
\begin{eqnarray}
\kappa_{b}&=&-\frac{1}{2}\dot{f}\Bigr|_{r=r_b}~,\label{surf-grav-bh}\\
\kappa_{c}&=&\frac{1}{2}\dot{f}\Bigr|_{r=r_c}~,\label{surf-grav-cosm}
\end{eqnarray}
where $f(r)\equiv e^{-2\beta}$. This quantity is directly related to the Hawking temperature,
\begin{equation}\label{T-hor}
T=\frac{\kappa}{2\pi}~,
\end{equation}
The second law of the thermodynamics of black holes states that the area of the surface of the event horizon can not decrease under any physical process \cite{Wald:1984rg}. Such entropy is calculated by\footnote{Since we are dealing with a perturbative geometry around a GR (plus cosmological constant) solution, the validity of Eqs.~\eqref{T-hor} and \eqref{S-hor} are, at least, a good approximation.}
\begin{equation}\label{S-hor}
S=\frac{A}{4G}\equiv\frac{\pi r^2}{G}~,
\end{equation}
where $r$ is the horizon radius and $A$ is the area of this surface. In this section, the aim is to compare our results with those from the standard literature on black holes and cosmological horizons \cite{Gibbons:1977mu,Wald:1984rg}.

The exact solution brings the similar results for the entropy to the de Sitter spacetime with a simple shift from $\tilde{\Lambda}^2$ to $\Upsilon$. In the region $r\gg 2GM$, it is remarkable that we connect both perturbative and exact entropies up to corrections in $\eta$. Of course, in the limit $\eta\rightarrow 0$, both entropies are matched.

\begin{figure}[htb]
	\centering
		\includegraphics[width=0.6\textwidth]{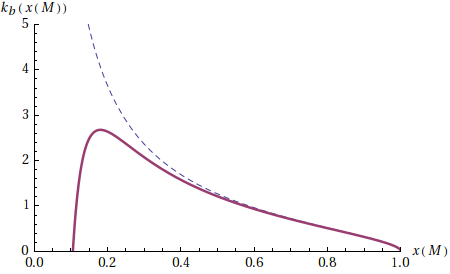}
		\caption{Surface gravity in the event horizon of the mass spherical distribution $M$. The dashed and thick curves stand for the standard surface gravity of a black hole and the perturbative surface gravity, respectively. For the thick curve we adopted $\eta=10^{-3}$.}
\label{graphkb}
\end{figure}

We calculate the surface gravities related to each horizon and we analyze their behaviours under the change of the mass $M$. In the perturbative case we have two horizons, $r_b$ and $r_c$, which are explicitly written in \eqref{HZs}. Let us start the analysis with the event horizon. From Eq.~\eqref{surf-grav-bh} we find
\begin{equation}\label{surf-grav-bh-a}
\kappa_b\approx -\frac{1}{3}\tilde{\Lambda}^2 r_{02} + \frac{2GM}{r_{02}^2} + \eta\left[-2GM\frac{r_{12}}{r_{02}^3} -\frac{1}{3}\tilde{\Lambda}^2r_{12} + \frac{GM}{r_{02}^2}-\frac{1}{3}\tilde{\Lambda}^2 r_{02} - \frac{12G^2M^2}{\tilde{\Lambda}^2}\frac{1}{r_{02}^5} \right]~.
\end{equation}
The behaviour of $\kappa_b$ are qualitatively displayed in Figure~\ref{graphkb}. We observe that the increasing of $M$ implies on the decreasing of $\kappa_b$. Accordingly, from Eq.~\eqref{T-hor}, it is clear that the temperature $T_b=\kappa_b/2\pi$ has the same behaviour as $\kappa_b$.

\begin{figure}[htb]
	\centering
		\includegraphics[width=0.6\textwidth]{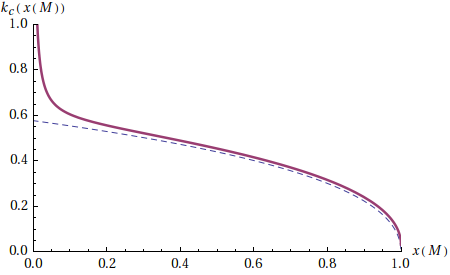}
		\caption{Surface gravity of the cosmological horizon.
		$r_c(x(M))$ is in units of $\tilde{\Lambda}^{-1}$, $3GM\tilde{\Lambda}\equiv x$ and $\eta=10^{-2}$.
		The dashed curve represents the surface gravity to the cosmological horizon obtained from the Schwarzschild-de Sitter geometry.
		The thick curve is the cosmological horizon with the correction from the quadratic term contribution.}
\label{graph-kc}
\end{figure}

For the cosmological horizon, Eq.~\eqref{surf-grav-cosm} provides
\begin{equation}\label{surf-grav-c}
\kappa_c\approx \frac{1}{3}\tilde{\Lambda}^2 r_{01} - \frac{2GM}{r_{01}^2} + \eta\left[2GM\frac{r_{11}}{r_{01}^3} +\frac{1}{3}\tilde{\Lambda}^2r_{11} - \frac{GM}{r_{01}^2}+\frac{1}{3}\tilde{\Lambda}^2 r_{01} + \frac{12G^2M^2}{\tilde{\Lambda}^2}\frac{1}{r_{01}^5} \right]~.
\end{equation}
The plot of the cosmological horizon surface gravity, as function of $M$, is displayed in Figure~\ref{graph-kc}. The behaviour of the surface gravity is essentially the same in both cases. According to Eq.~\eqref{T-hor}, the respective temperature $T_c=\kappa_c/2\pi$ have the same qualitative behaviour as the cosmological horizon.

\newpage

\begin{figure}[htb]
	\centering
		\includegraphics[width=0.6\textwidth]{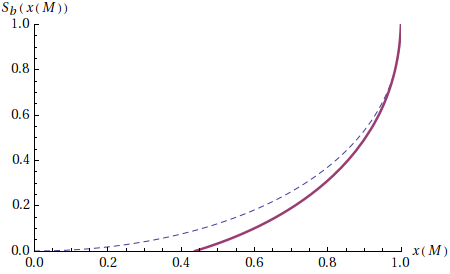}
		\caption{Entropy of the event horizon of the mass spherical distribution.
		$S_b(x(M))$ is in units of $\tilde{\Lambda}^{-1}$, $3GM\tilde{\Lambda}\equiv x$ and $\eta=10^{-2}$.
		The dashed curve represents the entropy to the standard black hole horizon obtained from the Schwarzschild-de Sitter geometry.
		The thick curve is the entropy with the correction from the quadratic term contribution.}
		\label{graph-sb}
\end{figure}

\begin{figure}[htb]
	\centering
		\includegraphics[width=0.6\textwidth]{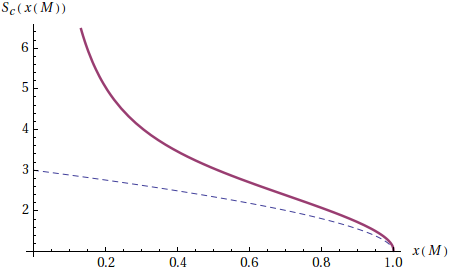}
		\caption{Entropy of the cosmological horizon.
		$S_c(x(M))$ is in units of $\tilde{\Lambda}^{-1}$, $3GM\tilde{\Lambda}\equiv x$ and $\eta=10^{-1}$.
		The dashed curve represents the entropy to the standard cosmological horizon obtained from the Schwarzschild-de Sitter geometry.
		The thick curve is the entropy with the correction from the quadratic term contribution.} \label{graph-sc}
\end{figure}

The entropies can be computed from Eq.~\eqref{S-hor}, providing\footnote{We will not display the full expressions to the entropies \eqref{S-rb} and \eqref{S-rc} due to their extension.}

\begin{eqnarray}
S_{b}&=&\frac{\pi r_b^2}{G}\approx \frac{\pi r_{01}^2}{G}\left[1+2\eta\left(\frac{r_{11}}{r_{01}}\right)^2\right]~,\label{S-rb}\\
S_c&=&\frac{\pi r_c^2}{G}\approx \frac{\pi r_{02}^2}{G}\left[1+2\eta\left(\frac{r_{12}}{r_{02}}\right)^2\right]~,\label{S-rc}
\end{eqnarray}
where $r_{11}$ and $r_{12}$ correspond to the corrections due to the quadratic curvature term of the horizons. It is clear from the plots that $S_b$ increases with the mass while $S_c$ decreases with the mass. We remark that entropy behaviours are in agreement with \cite{Gibbons:1977mu,Frolov:1981mz}, of course, with a small deviation due the perturbative approximation. The plots with the qualitative behaviour of such entropies are displayed in Figure~\ref{graph-sb} and Figure~\ref{graph-sc}.


Concerning the first law of thermodynamics, in the exact case, it is expected to be fully satisfied, matching with the similar results of \cite{Hawking:1974sw}, since we have a de Sitter-like spacetime and such spacetimes do not harm the first law of the thermodynamics.

In the perturbative case, the first law should also be satisfied. It certainly is satisfied at zeroth order because it is the Schwarzschild-de Sitter solution where the first law is satisfied \cite{Hawking:1974sw}. It is then expected that any correction arising from the perturbative expansion would be absorbed order by order. For instance, in the case of Lovelock-Lanczos theory there are corrections to the equations for temperature and entropy and such modifications are in agreement with the first law of thermodynamics \cite{Kothawala:2009kc}. Another example is the modifications on the temperature and entropy equations that occur in $f(R)$ theories of gravity \cite{Bamba:2010kf} and the first law is completely satisfied after all. Thus, in our case, the first law is satisfied at zeroth order and each correction in the temperature and entropy would generate a correction for the first law, making it valid at all orders in the perturbative series.

\section{Conclusions}\label{rmks}

In this work we found static and spherically symmetric solutions to a generalized gravity action in the EC formalism for vanishing torsion situations. The model, described by the action \eqref{ym-map-grav-obs}, is composed by the usual EH term, the cosmological constant term, a curvature squared term, and a quadratic torsion term. Our results are summarized here:
\begin{itemize}
\item The coupled system of differential equations \eqref{edo-t}-\eqref{edo-scalarcurv} is exactly solved. The result is a de Sitter spacetime described by \eqref{exact-sol-pm} with two possible effective cosmological constants described in \eqref{Ypm}. Both cases are compositions of the parameters $\tilde{\Lambda}$ and $\Lambda$. This result is important since the system \eqref{edo-t}-\eqref{edo-scalarcurv} is over-determined.

\item For the case where $\tilde{\Lambda}\ll\Lambda$, the effective cosmological constants are simplified (at first order) to Eq.~\eqref{Yappr}. Hence, two regimes are possible, one with a strong curvature and another with a weak curvature.

\item By treating the quadratic curvature term as a perturbation when compared do the rest of the terms, a perturbative solution is found. For that, Eq.~\eqref{edo-scalarcurv} can be neglected (this situation corresponds to the case where torsion is set to zero at the action level). Hence, the smaller system  system \eqref{edo-t}-\eqref{edo-theta-phi} is not over-determined. The solution \eqref{pert-sol-4th} is a deformation of the usual Schwarzschild-de Sitter solution.

\item The appropriate limits of the perturbative solution are all consistent. In particular, at the limit $r\gg 2GM$, the asymptotic spacetime is a de Sitter spacetime. The same spacetime is obtained from the exact solution for $\tilde{\Lambda}\ll\Lambda$. See \eqref{pert-ds-far}-\eqref{Y-notrunc}.

\item The usual singularity $r=0$ is present at the perturbative solution and no other singularities appear.

\item The horizons are obtained as perturbed deformations of the usual event and cosmological horizons of the Schwarzschild-de Sitter spacetime, namely \eqref{HZs}.

\item Moreover, the surface gravities and entropies associated to the horizons are also computed. The results are perturbations around the usual surface gravities and entropies of the Schwarzschild-de Sitter spacetime. See \eqref{surf-grav-bh-a}-\eqref{S-rc}.

\item It is also discussed the qualitative behaviour of the horizons, surface gravities and entropies as functions of the mass $M$ of the matter distribution. In all cases, the bigger the mass the better is the agreement between the perturbative solution and the Schwarzschild-de Sitter one. The deviations, \emph{i.e.}, the perturbative corrections, are stronger as the mass decreases.
\end{itemize}

At this point it is convenient to compare our perturbative solution \eqref{pert-sol-4th} with the one encountered by K.~S.~Stelle in \cite{Stelle:1977ry}. In that work, a general theory of gravity in the second order formalism with higher derivative terms (quadratic terms in the Riemann tensor, Ricci tensor and scalar curvature) is considered and a perturbative static and spherically symmetric solution is found. At first sight one could argue that our result should be just a particular case of Stelle`s result for a suitable choice of parameters. However, our result is strongly different. First of all, the result in \cite{Stelle:1977ry} is a perturbation around Minkowski spacetime while ours is a perturbation around the Schwarzschild-de Sitter solution. Second, our approach relies in solving a perturbed version of the field equations. In \cite{Stelle:1977ry} they propose a perturbed solution around Minkowski spacetime while keeping the field equations exact. As a consequence, their perturbed functions are different from ours. In fact, our perturbations consist on corrections of the Schwarzschild-de Sitter solution due to the quadratic curvature term while the solution in \cite{Stelle:1977ry} is a perturbation around the Minkowski spacetime, independently of the magnitude of the higher order derivative terms. Moreover, it is known that the field equations in the second order formalism are significantly different from the equations obtained in the first order formalism \cite{Exirifard:2007da,Borunda:2008kf,Capozziello:2010ih}, except for the special case of GR.

The study of static spherically symmetric solutions within action \eqref{ym-map-grav-obs} inside the matter distribution is currently under investigation for a perfect fluid distribution \cite{Silveira:2017xxx}. Moreover, solutions for spherically symmetric matter distributions in the presence of electromagnetic fields and rotation will be analyzed as well. As a perspective, the apparent horizons will be investigated in that work considering an energy-momentum tensor coupled with the first equation of motion of the induced gravity theory.

It is well-known, yet tricky, in the literature how to determine the necessary and sufficient conditions to prove the stability of spherically symmetric solutions. For instance, the Schwarzschild solution had its linear stability proved recently \cite{Dafermos:2016uzj,Chaverra:2012bh}. Concerning the stability of the solutions presented in the paper, the stability is currently under investigation. Another study of extreme relevance which is left for future investigation focuses on initial-value problem (if it is well-posed or not) of the modified theory of gravity considered in this paper. Both problems need extended study and deserves exclusive treatment by their own.

Finally, let us remark that the generalization of our results by considering non-vanishing torsion will also be studied. In this case, the system \eqref{eq:eq_mov1}-\eqref{eq:eq_mov2} has more degrees of freedom, a property that opens the possibility of new solutions than the ones here found.

\section*{Acknowledgements}

The Conselho Nacional de Desenvolvimento Cient\'ifico e Tecnol\'ogico (CNPq-Brazil), The Coordena\c c\~ao de Aperfei\c coamento de Pessoal de N\'ivel Superior (CAPES), the Pr\'o-Reitoria de Pesquisa, P\'os-Gradua\c c\~ao e Inova\c c\~ao (PROPPI-UFF) and Centro Brasileiro de Pesquisas F\'isicas are acknowledge for financial support. The authors are grateful to N\'estor Ortiz and Gustavo Pazzini de Brito for the clarifying discussions and the important comments on the manuscript.

\appendix

\section{On the convergence of the $u(r)$}\label{CSdemo}

Let us start the analysis by expanding the sum in $\ell$ in Eq.~\eqref{pert-sol-gen},
\begin{equation}\label{series1}
u(r)= r^2\sum_{k=0}^{\infty}\eta^k\mathcal{C}_{k1} + \frac{1}{r}\sum_{k=0}^{\infty}\eta^k\mathcal{C}_{k2} +\frac{1}{r^4}\sum_{k=1}^{\infty}\eta^k \mathcal{C}_{k3}+\frac{1}{r^7}\sum_{k=2}^{\infty}\eta^k\mathcal{C}_{k4}+\frac{1}{r^{10}}\sum_{k=3}^{\infty}\eta^k\mathcal{C}_{k5}+\frac{1}{r^{13}}\sum_{k=4}^{\infty}\eta^k\mathcal{C}_{k6}+\cdots ~.
\end{equation}
The series \eqref{series1} is more conveniently split in three pieces,
\begin{equation}\label{series2}
u(r)=u_{dS}(r) + u_S(r) +u_C(r)~,
\end{equation}
where
\begin{eqnarray}
u_{dS}(r)&=&r^2\sum_{k=0}^{\infty}\eta^k\mathcal{C}_{k1}~,\label{series2a}\\
u_S(r)&=&\frac{1}{r}\sum_{k=0}^{\infty}\eta^k\mathcal{C}_{k2}~,\label{series2b}\\
u_C(r)&=&\sum_{\ell=3}^{\infty}a_\ell r^{5-3\ell}\label{series2c}
\end{eqnarray}
with
\begin{equation}\label{series2d}
a_\ell=\sum_{k=\ell-2}^\infty\eta^k\mathcal{C}_{k\ell}~.
\end{equation}
The absolute and uniform convergence of Eq.~\eqref{series2} is ensured if, and only if, \eqref{series2a}, \eqref{series2b} and \eqref{series2c} are convergent by themselves. Thus, we shall deal with each term separately as follows.

\subsection{de Sitter series}

The series represented by $u_{dS}$, which describes the de Sitter sector of the perturbative solution, can be expressed as
\begin{equation}
u_{dS}(r)=\tilde{\Upsilon}_mr^2,
\end{equation}
with
\begin{equation}\label{series3}
\tilde{\Upsilon}_m=\frac{\tilde{\Lambda}^2}{3}\left(1+\eta+2\eta^2+5\eta^3+14\eta^4+\cdots\right)\equiv \sum_{w=0}^\infty b_w\equiv\frac{\Lambda^2}{3}\left(1-\sqrt{1-4\eta}\right)~,
\end{equation}
where $\eta=\tilde{\Lambda}^2/(2\Lambda^2)$, $b_w=(\tilde{\Lambda}^2/3)\eta^w c_w$ and $c_w$ are the Catalan numbers -- See Eq.~\eqref{catalanNum}. The Eq.~\eqref{series3} elucidate that result in the perturbative solution matches with the exact one in the limit $r\rightarrow\infty$ as such as aforementioned in Sect.~\ref{pertsol}. Moreover, it means that the series can be exactly re-summed if it converges. The condition for the convergence of \eqref{series3} is obtained from the D'Alembert ratio criterion \cite{arfken2008mathematical} by means of 
\begin{equation}
\lim_{w\to\infty} \left|\frac{b_{w+1}}{b_w}\right|=\lim_{w\to\infty}\frac{\frac{\tilde{\Lambda}^2}{3}\eta^{w+1}\left[\frac{(2w+2)!}{(w+2)!(w+1)!}\right]}{\frac{\tilde{\Lambda}^2}{3}\eta^w\left[\frac{(2w)!}{(w+1)!w!}\right]}=4\eta<1~,
\end{equation}
therefore, $u_{dS}(r)$ is absolutely convergent under the D'Alembert ratio criterion if, and only if,
\begin{equation}
\eta<\frac{1}{4}\;.\label{cond1}
\end{equation}

\subsection{Schwarzschild series}

The series $u_S(r)$ contains all the integration constants $\mathcal{C}_{k\ell}$. As aforementioned in Sect.~\ref{pertsol}, it is possible to  re-sum these constants to rewrite this series in a compact form by defining an effective mass $\tilde{M}$, \emph{i.e.}
\begin{equation}
u_S(r)= \frac{2G\tilde{M}}{r},
\end{equation}
where $2G\tilde{M}=2GM+\eta \mathcal{C}_{12}+ \eta^2 \mathcal{C}_{22} + \eta^3 \mathcal{C}_{32}+\cdots$. Employing the D'Alembert ratio criterion again, we find, 
\begin{equation}\lim_{k\to\infty}\left|\frac{\eta^{k+1}\mathcal{C}_{(k+1)2}}{\eta^k \mathcal{C}_{k2}}\right|<1,
\end{equation}
from which we obtain the second condition for convergence
\begin{equation}\label{ratio00}
\left|\frac{\mathcal{C}_{(k+1)2}}{\mathcal{C}_{k2}}\right|<\frac{1}{\eta}~,
\end{equation}
where we used the fact that the constants $\mathcal{C}_{k2}$ are independent of $k$. Under the above condition, the series $u_S(r)$ is absolutely convergent.

\subsection{Higher order series}

At last, we analyze the convergence of $u_C(r)$, which carries all higher order term in $r^{-n}$ for $n\ge4$. First, we employ the D'Alembert ratio criterion for Eq.~\eqref{series2d}, providing
\begin{equation}\label{ratio01}
\lim_{k\to\infty} \left|\frac{\eta^{k+1}\mathcal{C}_{\left(k+1\right)\ell}}{\eta^k\mathcal{C}_{k\ell}}\right|< 1~.
\end{equation}
The constants $\mathcal{C}_{(k+1)\ell}$ and $\mathcal{C}_{k\ell}$ are independent of $k$, $\forall \ell\geq 3$. Thus, \eqref{ratio01} simplifies to
\begin{equation}
|\frac{\mathcal{C}_{\left(k+1\right)\ell}}{\mathcal{C}_{k\ell}}|<\frac{1}{\eta}\;,~\forall k\in[0,\infty) ~~\mathrm{and}~~\forall \ell\in[3,\infty)\;,\label{ratio02}
\end{equation}
just like \eqref{ratio00}. In fact, \eqref{ratio02} generalizes \eqref{ratio00} for the range $k\in[0,\infty)$ and $\ell\in[2,\infty)$. The constraint \eqref{ratio02} guarantees that each $a_\ell$ absolutely converges.\\
Now, to see if $u_C(r)$ converges as a whole, we employ Abel's uniform convergence criterion \cite{arfken2008mathematical}, which states:\\\\
\textit{Let $\{w_n(r)\}$ be a sequence of functions. If
\begin{itemize}
\item[(i)] $w_n(r)$ can be written as $w_n(r)=a_n f_n(r)$,
\item[(ii)] $\sum a_n$ is convergent,
\item [(iii)] $f_n(r)$ is a monotonic decreasing sequence, \emph{i.e.}, $f_{n+1}(r)\leq f_n(r)$,
\item[(iv)] $f_n(r)$ is bounded in some region, \emph{i.e.}, $0\leq f_n(r)\leq \mathcal{F}$, $\forall r\in [a,b]$,
\end{itemize}}
\noindent\textit{Then, $w(r)=\sum a_nf_n(r)$ is said to absolutely convergent.}\\\\
Hence, we need to deal with each item above stated:
\begin{itemize}
\item[(i)] The first condition is satisfied since we can write the expression \eqref{series2c} in the form stated in (i) by defining $w_\ell=a_\ell f_\ell(r)$, where $f_\ell(r)=r^{5-3\ell}$.
\item[(ii)] $\sum_\ell a_\ell$ is absolutely convergent if (the D'Alembert ratio criterion is employed once again),
\begin{equation}
\lim_{\ell\rightarrow\infty}\left|\frac{a_{\ell+1}}{a_\ell} \right|<1~,
\end{equation}
thus, the constants $\mathcal{C}_{k\ell}$ must also obey the condition
\begin{equation}\label{ratio03}
\frac{\left|\sum_{k=\ell-1}^\infty \eta^k\mathcal{C}_{k(\ell+1)}\right|}{\left|\sum_{k=\ell-2}^{\infty} \eta^k\mathcal{C}_{k\ell}\right|}<1~.
\end{equation}
\item[(iii)] Since $r^{5-3(\ell+1)}<r^{5-3\ell}\Rightarrow f_{\ell+1}(r)<f_\ell(r)$, then $f_\ell(r)$ is monotonic decreasing. The third item is valid.
\item[(iv)] $0\leq f_\ell(r)\leq \mathcal{F}$ is also verified. To see this, we take a generic interval $r\in[a,b]$, providing, for a generic $f_\ell(r)$, $b^{5-3\ell}\leq f_\ell(r)\leq a^{5-3\ell}$, which is bounded for all intervals $[a,b]$ in $r\in[2GM,\infty)$. Even for the extreme case $r\rightarrow\infty$, $0\leq f_\ell(r)\leq  a^{5-3\ell}$, the functions $f_\ell(r)$ are bounded.\\\\
Thus, $u_C(r)$ converges absolutely and uniformly, provided condition \eqref{ratio03}.
\end{itemize}
Under the criteria established above (see \eqref{convergence00}), the series $u(r)$ is then convergent.

\section{A similar perturbative solution}\label{EDO9}

The aim of this appendix is to show that the solution \eqref{pert-sol-gen} is consistent with the differential equation \eqref{edo-theta-phi}. First, we write Eq.~\eqref{edo-theta-phi} (already assuming $\alpha+\beta=0$) as
\begin{equation}\label{edo9}
\frac{1}{4}\eta\left(2\frac{\dot{z}^2}{r^2}+\ddot{z}^2\right)-\lambda\left(\frac{\ddot{z}}{2}+\frac{\dot{z}}{r}+3\lambda\right)=0~,
\end{equation}
where $z\equiv z(r)=1-e^{-2\beta}$ and $\eta=\sigma\lambda$. The perturbative solution has the form
\begin{equation}\label{zPert}
z=\sum_{p=0}^\infty \eta^p z_p(r)~.
\end{equation}
Putting Eq.~\eqref{zPert} into Eq.~\eqref{edo9}, we obtain the following system of iterated differential equations:
\begin{eqnarray}
3\lambda+\frac{\dot{z}_0}{r}+\frac{1}{2} \ddot{z}_0=0~,\\
\frac{1}{4} \left(\frac{2\dot{z}_0^2}{r^2}+\ddot{z}_0^2\right)-\lambda\left(\frac{\dot{z}_1}{r}+\frac{1}{2} \ddot{z}_1\right)=0~,\\
\frac{1}{4}\left(\frac{4 \dot{z}_0 \dot{z}_1}{r^2}+2\ddot{z}_0\ddot{z}_1\right)-\lambda\left(\frac{\dot{z}_2}{r}+\frac{1}{2}\ddot{z}_2\right)=0~,\\
\frac{1}{4}\left[\frac{2\left(\dot{z}_1^2+2 \dot{z}_0\dot{z}_2\right)}{r^2}+\ddot{z}_1^2+2\ddot{z}_0\ddot{z}_2\right]-\lambda\left(\frac{\dot{z}_3}{r}+\frac{1}{2}\ddot{z}_3\right)=0~,\\
\frac{1}{4}\left[\frac{2\left(2\dot{z}_1 \dot{z}_2+2\dot{z}_0 \dot{z}_3\right)}{r^2}+2 \ddot{z}_1 \ddot{z}_2+2 \ddot{z}_0 \ddot{z}_3\right]-\lambda\left(\frac{\dot{z}_4}{r}+\frac{1}{2} \ddot{z}_4\right)=0~,\\
\vdots\nonumber
\end{eqnarray}
The zeroth order at $\eta$, \emph{i.e.}, $p=0$, has the solution
\begin{equation}
z_0=-\lambda r^2+\frac{\mathcal{D}_{02}}{r}+\mathcal{E}_0~,
\end{equation}
where $\mathcal{D}_{02}$ and $\mathcal{E}_0$ are constants of integration. Assuming the Newtonian limit, we get $\mathcal{D}_{02}=2GM$. To match with a genuine Schwarzschild-de Sitter solution in this order at $\eta$ it is necessary to fix $\mathcal{E}_0=0$. Recalling that $\lambda=-\tilde{\Lambda}^2/3$, we get
\begin{equation}
\left(e^{-2\beta}\right)_0=1-z_0=1-\frac{\tilde{\Lambda}^2}{3}r^2-\frac{2GM}{r}~
\end{equation}
with $\left(e^{-2\beta}\right)_0$ indicating the zeroth order of the solution. Next, we solve the first order by plugging the zeroth order solution into the first order equation. The result is given by
\begin{equation}
z_1=\mathcal{D}_{11} r^2+\frac{\mathcal{D}_{12}}{r}+\frac{\mathcal{D}_{13}}{r^4}+\mathcal{E}_1~,
\end{equation}
where $\mathcal{D}_{12}$ and $\mathcal{E}_1$ are constants of integration. Again, we must set $\mathcal{E}_1=0$.

For the next orders, we proceed recursively. The first compatibility condition is that, at all orders, we must demand that $\mathcal{E}_p=0$. To illustrate the other conditions for the remaining integration constants we work out the solution up to the fourth order, providing
\begin{eqnarray}\label{pertsol-edo9}
e^{-2\beta}&=& 1-\frac{2GM}{r}-\frac{\tilde{\Lambda}^2}{3}r^2-\eta\left(\frac{\mathcal{D}_{12}}{r}+\mathcal{D}_{11}r^2+\frac{\mathcal{D}_{13}}{r^4}\right)- \eta^2\left(\frac{\mathcal{D}_{22}}{r}+\mathcal{D}_{21}r^2+\frac{\mathcal{D}_{23}}{r^4}+\frac{\mathcal{D}_{24}}{r^7}\right)\nonumber\\
&-&\eta^3\left(\frac{\mathcal{D}_{32}}{r}+\mathcal{D}_{31}r^2+\frac{\mathcal{D}_{32}}{r^4}+\frac{\mathcal{D}_{34}}{r^7}+\frac{\mathcal{D}_{35}}{r^{10}}\right) - \eta^4\left(\frac{\mathcal{D}_{42}}{r}+\mathcal{D}_{41}r^2+\frac{\mathcal{D}_{43}}{r^4}+\frac{\mathcal{D}_{44}}{r^7}+\frac{\mathcal{D}_{45}}{r^{10}}+\frac{\mathcal{D}_{46}}{r^{13}}\right)+\ldots~,\nonumber\\
\end{eqnarray}
which can be generalized to
\begin{equation}
e^{-2\beta}=1-\sum_{p=0}^\infty\eta^p\sum_{q=1}^{p+2}\mathcal{D}_{pq}~.
\end{equation}
The constants $\mathcal{D}_{pq}$'s are listed in a matrix form as
\begin{equation}
\mathcal{D}_{pq}\equiv\left(\begin{array}{cccccc}\label{DIs}
\frac{\tilde{\Lambda}^2}{3} & 2GM  &  & & & \\
\frac{\tilde{\Lambda}^2}{3} & \mathcal{D}_{12}  & \frac{3G^2M^2}{\tilde{\Lambda}^2} & & & \\
2\frac{\tilde{\Lambda}^2}{3} & \mathcal{D}_{22} & \frac{3GM}{\tilde{\Lambda}^2}\Phi_1 & \frac{144G^3M^3}{7\tilde{\Lambda}^4} &  & \\
5\frac{\tilde{\Lambda}^2}{3} & \mathcal{D}_{32} & \frac{3}{4\tilde{\Lambda}^2}\Phi_2 & \frac{72G^2M^2}{7\tilde{\Lambda}^4}\Phi_3 & \frac{1188G^4M^4}{5\tilde{\Lambda}^6} & \\
14\frac{\tilde{\Lambda}^2}{3} & \mathcal{D}_{42}  & \frac{3}{2\tilde{\Lambda}^2}\Phi_4 & \frac{36GM}{7\tilde{\Lambda}^4}\Phi_5 & \frac{2376G^3M^3}{5\tilde{\Lambda}^6}\Phi_6 & \frac{46656G^5M^5}{\tilde{\Lambda}^8}
\end{array}\right)
\end{equation}
where $p\in\{0,1,2,3,\cdots\}$ and $q\in\{1,\cdots,p+2\}$ indicate the line and the column, respectively. The constants $\Phi_i$ are
\begin{eqnarray}
\Phi_1 &=&\mathcal{D}_{12}+2GM ~,\nonumber\\
\Phi_2 &=&\mathcal{D}_{12}\left(\mathcal{D}_{12}+8GM\right)+4GM\left(\mathcal{D}_{22}+6GM\right)~,\nonumber\\
\Phi_3 &=& 3\mathcal{D}_{12}+8GM~,\nonumber\\
\Phi_4 &=& \mathcal{D}_{12}^2+\mathcal{D}_{12}\left(\mathcal{D}_{22}+12GM\right)+2GM\left(2\mathcal{D}_{22} + \mathcal{D}_{32}+20GM\right)~,\nonumber\\
\Phi_5 &=& 3\mathcal{D}_{12}^2+24GM\mathcal{D}_{12}+2GM\left(3\mathcal{D}_{22} + 32GM\right)~,\nonumber\\
\Phi_6 &=& \mathcal{D}_{12}+3GM\nonumber~.
\end{eqnarray}
The redefinition of the constants obeys the rule $\mathcal{D}_{pq}\equiv f_{(pq)}\mathcal{C}_{k\ell}$ with $p=k$, $q=\ell$ and $f_{(pq)}$ (no summation is assumed in the relation) as a proportionality factor. We recall that the $\mathcal{C}_{k\ell}$ coefficients are casted in \eqref{CIs}. As an example, let us consider two integration constants, namely $\mathcal{D}_{01}$ and $\mathcal{D}_{13}$. For the former, we have that $\mathcal{D}_{01}\equiv f_{01}\mathcal{C}_{01}$ which implies $f_{01}=1$; For the latter we get that $\mathcal{D}_{13}\equiv f_{13}\mathcal{C}_{13}$ which implies $f_{13}=2$; and so on. We remark that some factors may depend on the arbitrary integration constants, the Newtonian constant $G$, the mass $M$ and the observational cosmological constant $\tilde{\Lambda}^2$.

Finally, we settled sufficient and necessary conditions to demonstrate the equivalence between the perturbative solution of Eq.~\eqref{edo-theta-phi} and the one of Eq.~\eqref{edo-t}, provided $\alpha=-\beta$.

\section{The Newtonian limit from the perspective of the effective potential}\label{NewtonLIM}

It is well known that Einstein's theory lead us to a corrected Newtonian potential. The corrected potential is obtained by considering a massive particle in a radial movement in the equatorial plane and using spherical symmetry. The Einsteinian effective potential obtained from General Relativity is
\begin{equation}\label{EinPot}
\Phi_{Einstein}=-\frac{GM}{r} + \frac{L^2}{2r^2} - \frac{GML^2}{r^3}~,
\end{equation}
where the first term is the usual Newtonian potential, the second term is the centrifugal potential and, the last term corresponds to the correction from Einstein's theory. $L$ is the particle's angular momentum per mass of the particle. When $r$ is large the first term in \eqref{EinPot} dominates and the Newtonian limit is ensured.

Following the well-established procedure \cite{Carroll:2004st,Ryder:2009zz,Hartle:2003yu,Misner:1974qy}, the effective potential can be computed from Eq.~\eqref{pert-sol-gen},
\begin{equation}\label{EffPot}
\Phi_\textrm{IG}=\frac{1}{2}\left[\frac{L^2}{r^2}-\sum_{k=0}^{\infty}\eta^k\sum_{\ell=1}^{k+2}
\mathcal{C}_{k\ell}r^{3(1-\ell)}\left(L^2+r^2\right)\right]~.
\end{equation}
Since the solution is perturbative, it is safe to look only at the first order,
\begin{equation}\label{EffPotIG}
\Phi_\textrm{IG}^{(1)}=\frac{1}{2}\left[ \frac{L^2}{r^2} -\mathcal{C}_{01}\left(L^2+r^2\right) - \mathcal{C}_{02}\left(\frac{L^2}{r^3}+\frac{1}{r}\right) - \eta\mathcal{C}_{11}\left(L^2+r^2\right)-\eta\mathcal{C}_{12}\left(\frac{L^2}{r^3}+\frac{1}{r}\right)-\eta\mathcal{C}_{13}\left(\frac{L^2}{r^6}+\frac{1}{r^4}\right)~\right].
\end{equation}
To check if \eqref{EffPotIG} respects the Newtonian limit, we can take $\eta\to 0$. Moreover, the observational cosmological constant $\mathcal{C}_{01}=\tilde{\Lambda}^2/3$ can also be neglected due to its small value. Thus, we are left only with terms proportional to powers of $1/r$.  Therefore, Eq.~\eqref{EffPotIG} is simplified to
\begin{equation}\label{EffPotIG_simp}
\Phi_\textrm{IG}^{(0)}=\frac{1}{2}\left[\frac{L^2}{r^2}-\mathcal{C}_{02}\left(\frac{1}{r}+\frac{L^2}{r^3}\right)\right]~.
\end{equation}
Comparing Eqs.~\eqref{EffPotIG_simp} and \eqref{EinPot}, the Einsteinian effective potential, as well as the Newtonian one, is recovered if, and only if, $\mathcal{C}_{02}=2GM$. Hence, the Newtonian limit is guaranteed. This result is expected since the our perturbative solution is a perturbation around the Schwarzschild-de Sitter solution with a small cosmological constant. However, a bit more care is required at the perturbation term due to the arbitrary constants $\mathcal{C}_{12}$ and $\mathcal{C}_{13}$. So they can influence the effect of $\eta$.

Let us start with $\mathcal{C}_{12}$. From the convergence's criteria \eqref{convergence00}, we see that the arbitrary integration constants obey a bound related to the $\eta$ parameter. Thus, all the $\eta$ values adopted to plot the induced gravity effective potential must attend the exigence that $\eta<0.25$. The next criterion in \eqref{convergence00} establishes that $\eta\mathcal{C}_{12}<\mathcal{C}_{02}=2GM$. For instance, the assumption  $\mathcal{C}_{12}=2GM$ attends these criteria. To parameterize the possible values for $\mathcal{C}_{12}$ we have freedom to write $\mathcal{C}_{12}=a\times 2GM$, where $a\in(0,4)$, according to the above analysis. In this range, no meaningful modifications in the shape of $\Phi_{IG}^{(1)}$ are detected. Moreover, it is important to mention that the classical tests of General Relativity provides bounds to $\mathcal{C}_{12}$ which was already computed in \cite{daRocha:2017tiz}. The most restrictive bound is related to the perihelion precession of Mercury, which provides $|\eta\mathcal{C}_{12}|<\left(4.3\pm 3.1\right)\times 10^{28} kg$ (for $G=1$). 

The analysis for $\mathcal{C}_{13}$, which corrects the cosmological constant follows the same line of reasoning and provides $\eta\mathcal{C}_{13}<\mathcal{C}_{03}=\tilde{\Lambda}^2/3$. This criterion is quite stronger due to the small of the observational cosmological constant. Nevertheless, it ensures that the correction on the cosmological constant would be irrelevant for observational proposals.

Figure~\ref{PotEffs} shows the influence of the extra terms (terms proportional to $\eta$ and the observational cosmological constant) appearing in Eq.~\eqref{EffPotIG}. We have chosen six different values for $\eta$ to clarify the contribution of the perturbative solution displayed in Eq.~\eqref{pert-sol-gen} (at first order) to the effective potential related to the induced gravity theory.

\begin{figure}[htb]\label{PotEffs}
\includegraphics[width=\textwidth]{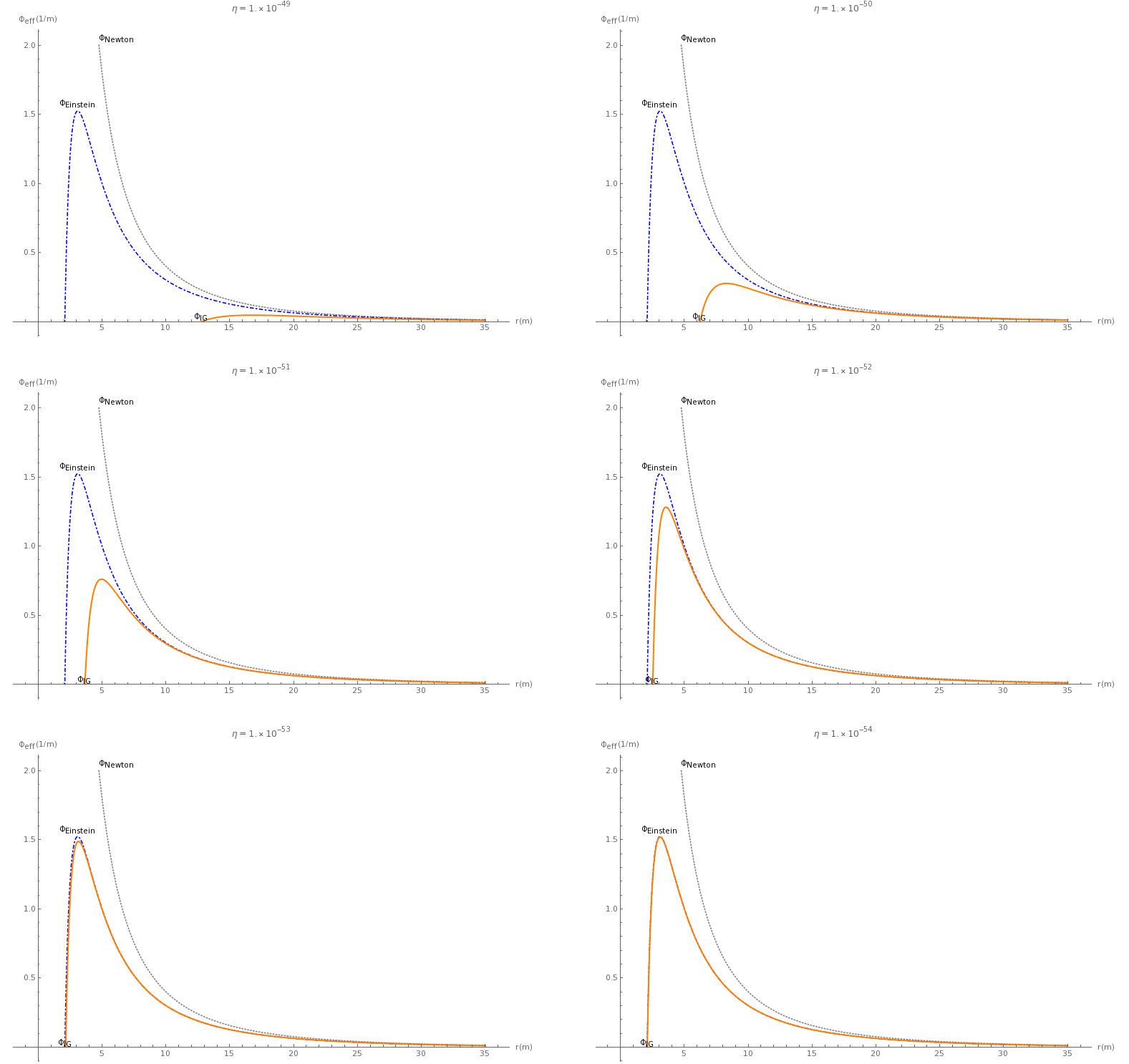}
\caption{The effective potential -- truncated at first order --  from the induced gravity $\Phi_\textrm{IG}^{(1)}$, the Einsteinian effective potential $\Phi_\textrm{Einstein}$ and the Newtonian one $\Phi_\textrm{Newton}$, respectively, the orange thick line, the blue dot-dashed line and the gray dotted line. The values used in each plot are $L=10$, $\tilde{\Lambda}^2=10^{-52}m^{-2}$, $GM=1$, $\mathcal{C}_{02}=\mathcal{C}_{12}=2GM$. In the limit for large values of $r$, the induced gravity effective potential assimilates the Newtonian effective potential as expected.}\label{PotEffs}
\end{figure}

\newpage
\bibliographystyle{apsrev4-1}
\bibliography{Bibliography}

\begin{thebibliography}{54}%
\makeatletter
\providecommand \@ifxundefined [1]{%
 \@ifx{#1\undefined}
}%
\providecommand \@ifnum [1]{%
 \ifnum #1\expandafter \@firstoftwo
 \else \expandafter \@secondoftwo
 \fi
}%
\providecommand \@ifx [1]{%
 \ifx #1\expandafter \@firstoftwo
 \else \expandafter \@secondoftwo
 \fi
}%
\providecommand \natexlab [1]{#1}%
\providecommand \enquote  [1]{``#1''}%
\providecommand \bibnamefont  [1]{#1}%
\providecommand \bibfnamefont [1]{#1}%
\providecommand \citenamefont [1]{#1}%
\providecommand \href@noop [0]{\@secondoftwo}%
\providecommand \href [0]{\begingroup \@sanitize@url \@href}%
\providecommand \@href[1]{\@@startlink{#1}\@@href}%
\providecommand \@@href[1]{\endgroup#1\@@endlink}%
\providecommand \@sanitize@url [0]{\catcode `\\12\catcode `\$12\catcode
  `\&12\catcode `\#12\catcode `\^12\catcode `\_12\catcode `\%12\relax}%
\providecommand \@@startlink[1]{}%
\providecommand \@@endlink[0]{}%
\providecommand \url  [0]{\begingroup\@sanitize@url \@url }%
\providecommand \@url [1]{\endgroup\@href {#1}{\urlprefix }}%
\providecommand \urlprefix  [0]{URL }%
\providecommand \Eprint [0]{\href }%
\providecommand \doibase [0]{http://dx.doi.org/}%
\providecommand \selectlanguage [0]{\@gobble}%
\providecommand \bibinfo  [0]{\@secondoftwo}%
\providecommand \bibfield  [0]{\@secondoftwo}%
\providecommand \translation [1]{[#1]}%
\providecommand \BibitemOpen [0]{}%
\providecommand \bibitemStop [0]{}%
\providecommand \bibitemNoStop [0]{.\EOS\space}%
\providecommand \EOS [0]{\spacefactor3000\relax}%
\providecommand \BibitemShut  [1]{\csname bibitem#1\endcsname}%
\let\auto@bib@innerbib\@empty
\bibitem [{\citenamefont {Einstein}(1916)}]{Einstein:1916vd}%
  \BibitemOpen
  \bibfield  {author} {\bibinfo {author} {\bibfnamefont {A.}~\bibnamefont
  {Einstein}},\ }\href@noop {} {\bibfield  {journal} {\bibinfo  {journal}
  {Annalen Phys.}\ }\textbf {\bibinfo {volume} {49}},\ \bibinfo {pages} {769}
  (\bibinfo {year} {1916})},\ \bibinfo {note} {[Annalen
  Phys.14,517(2005)]}\BibitemShut {NoStop}%
\bibitem [{\citenamefont {Utiyama}(1956)}]{Utiyama:1956sy}%
  \BibitemOpen
  \bibfield  {author} {\bibinfo {author} {\bibfnamefont {R.}~\bibnamefont
  {Utiyama}},\ }\href@noop {} {\bibfield  {journal} {\bibinfo  {journal} {Phys.
  Rev.}\ }\textbf {\bibinfo {volume} {101}},\ \bibinfo {pages} {1597} (\bibinfo
  {year} {1956})}\BibitemShut {NoStop}%
\bibitem [{\citenamefont {Kibble}(1961)}]{Kibble:1961ba}%
  \BibitemOpen
  \bibfield  {author} {\bibinfo {author} {\bibfnamefont {T.~W.~B.}\
  \bibnamefont {Kibble}},\ }\href@noop {} {\bibfield  {journal} {\bibinfo
  {journal} {J. Math. Phys.}\ }\textbf {\bibinfo {volume} {2}},\ \bibinfo
  {pages} {212} (\bibinfo {year} {1961})}\BibitemShut {NoStop}%
\bibitem [{\citenamefont {Sciama}(1964)}]{Sciama:1964wt}%
  \BibitemOpen
  \bibfield  {author} {\bibinfo {author} {\bibfnamefont {D.~W.}\ \bibnamefont
  {Sciama}},\ }\href@noop {} {\bibfield  {journal} {\bibinfo  {journal} {Rev.
  Mod. Phys.}\ }\textbf {\bibinfo {volume} {36}},\ \bibinfo {pages} {463}
  (\bibinfo {year} {1964})}\BibitemShut {NoStop}%
\bibitem [{\citenamefont {Zanelli}(2005)}]{Zanelli:2005sa}%
  \BibitemOpen
  \bibfield  {author} {\bibinfo {author} {\bibfnamefont {J.}~\bibnamefont
  {Zanelli}},\ }in\ \href@noop {} {\emph {\bibinfo {booktitle} {{Proceedings,
  7th Mexican Workshop on Particles and Fields (MWPF 1999)}}}}\ (\bibinfo
  {year} {2005})\ \Eprint {http://arxiv.org/abs/hep-th/0502193}
  {hep-th/0502193} \BibitemShut {NoStop}%
\bibitem [{\citenamefont {Bergmann}\ and\ \citenamefont
  {De~Sabbata}(1980)}]{Bergmann:1980wt}%
  \BibitemOpen
  \bibfield  {author} {\bibinfo {author} {\bibfnamefont {P.~G.}\ \bibnamefont
  {Bergmann}}\ and\ \bibinfo {author} {\bibfnamefont {V.}~\bibnamefont
  {De~Sabbata}},\ }\href@noop {} {\bibfield  {journal} {\bibinfo  {journal}
  {NATO Sci. Ser. B}\ }\textbf {\bibinfo {volume} {58}},\ \bibinfo {pages}
  {pp.1} (\bibinfo {year} {1980})}\BibitemShut {NoStop}%
\bibitem [{\citenamefont {Ferraris}\ \emph {et~al.}(1982)\citenamefont
  {Ferraris}, \citenamefont {Francaviglia},\ and\ \citenamefont
  {Reina}}]{Ferraris1982}%
  \BibitemOpen
  \bibfield  {author} {\bibinfo {author} {\bibfnamefont {M.}~\bibnamefont
  {Ferraris}}, \bibinfo {author} {\bibfnamefont {M.}~\bibnamefont
  {Francaviglia}}, \ and\ \bibinfo {author} {\bibfnamefont {C.}~\bibnamefont
  {Reina}},\ }\href {\doibase 10.1007/BF00756060} {\bibfield  {journal}
  {\bibinfo  {journal} {General Relativity and Gravitation}\ }\textbf {\bibinfo
  {volume} {14}},\ \bibinfo {pages} {243} (\bibinfo {year} {1982})}\BibitemShut
  {NoStop}%
\bibitem [{\citenamefont {De~Sabbata}\ and\ \citenamefont
  {Gasperini}(1986)}]{DeSabbata:1986sv}%
  \BibitemOpen
  \bibfield  {author} {\bibinfo {author} {\bibfnamefont {V.}~\bibnamefont
  {De~Sabbata}}\ and\ \bibinfo {author} {\bibfnamefont {M.}~\bibnamefont
  {Gasperini}},\ }\href@noop {} {\emph {\bibinfo {title} {Introduction to
  Gravitation}}}\ (\bibinfo  {publisher} {World Scientific},\ \bibinfo {year}
  {1986})\BibitemShut {NoStop}%
\bibitem [{\citenamefont {Stelle}(1978)}]{Stelle:1977ry}%
  \BibitemOpen
  \bibfield  {author} {\bibinfo {author} {\bibfnamefont {K.~S.}\ \bibnamefont
  {Stelle}},\ }\href@noop {} {\bibfield  {journal} {\bibinfo  {journal} {Gen.
  Rel. Grav.}\ }\textbf {\bibinfo {volume} {9}},\ \bibinfo {pages} {353}
  (\bibinfo {year} {1978})}\BibitemShut {NoStop}%
\bibitem [{\citenamefont {Seifert}(2007)}]{Seifert:2007fr}%
  \BibitemOpen
  \bibfield  {author} {\bibinfo {author} {\bibfnamefont {M.~D.}\ \bibnamefont
  {Seifert}},\ }\href {\doibase 10.1103/PhysRevD.76.064002} {\bibfield
  {journal} {\bibinfo  {journal} {Phys. Rev.}\ }\textbf {\bibinfo {volume}
  {D76}},\ \bibinfo {pages} {064002} (\bibinfo {year} {2007})},\ \Eprint
  {http://arxiv.org/abs/gr-qc/0703060} {arXiv:gr-qc/0703060 [gr-qc]}
  \BibitemShut {NoStop}%
\bibitem [{\citenamefont {Sebastiani}\ and\ \citenamefont
  {Zerbini}(2011)}]{Sebastiani:2010kv}%
  \BibitemOpen
  \bibfield  {author} {\bibinfo {author} {\bibfnamefont {L.}~\bibnamefont
  {Sebastiani}}\ and\ \bibinfo {author} {\bibfnamefont {S.}~\bibnamefont
  {Zerbini}},\ }\href {\doibase 10.1140/epjc/s10052-011-1591-8} {\bibfield
  {journal} {\bibinfo  {journal} {Eur. Phys. J.}\ }\textbf {\bibinfo {volume}
  {C71}},\ \bibinfo {pages} {1591} (\bibinfo {year} {2011})},\ \Eprint
  {http://arxiv.org/abs/1012.5230} {arXiv:1012.5230 [gr-qc]} \BibitemShut
  {NoStop}%
\bibitem [{\citenamefont {Lu}\ \emph {et~al.}(2015)\citenamefont {Lu},
  \citenamefont {Perkins}, \citenamefont {Pope},\ and\ \citenamefont
  {Stelle}}]{Lu:2015psa}%
  \BibitemOpen
  \bibfield  {author} {\bibinfo {author} {\bibfnamefont {H.}~\bibnamefont
  {Lu}}, \bibinfo {author} {\bibfnamefont {A.}~\bibnamefont {Perkins}},
  \bibinfo {author} {\bibfnamefont {C.~N.}\ \bibnamefont {Pope}}, \ and\
  \bibinfo {author} {\bibfnamefont {K.~S.}\ \bibnamefont {Stelle}},\ }\href
  {\doibase 10.1103/PhysRevD.92.124019} {\bibfield  {journal} {\bibinfo
  {journal} {Phys. Rev.}\ }\textbf {\bibinfo {volume} {D92}},\ \bibinfo {pages}
  {124019} (\bibinfo {year} {2015})},\ \Eprint
  {http://arxiv.org/abs/1508.00010} {arXiv:1508.00010 [hep-th]} \BibitemShut
  {NoStop}%
\bibitem [{\citenamefont {Nojiri}\ and\ \citenamefont
  {Odintsov}(2013)}]{Nojiri:2013su}%
  \BibitemOpen
  \bibfield  {author} {\bibinfo {author} {\bibfnamefont {S.}~\bibnamefont
  {Nojiri}}\ and\ \bibinfo {author} {\bibfnamefont {S.~D.}\ \bibnamefont
  {Odintsov}},\ }\href {\doibase 10.1088/0264-9381/30/12/125003} {\bibfield
  {journal} {\bibinfo  {journal} {Class. Quant. Grav.}\ }\textbf {\bibinfo
  {volume} {30}},\ \bibinfo {pages} {125003} (\bibinfo {year} {2013})},\
  \Eprint {http://arxiv.org/abs/1301.2775} {arXiv:1301.2775 [hep-th]}
  \BibitemShut {NoStop}%
\bibitem [{\citenamefont {Cognola}\ \emph
  {et~al.}(2015{\natexlab{a}})\citenamefont {Cognola}, \citenamefont
  {Rinaldi},\ and\ \citenamefont {Vanzo}}]{Cognola:2015uva}%
  \BibitemOpen
  \bibfield  {author} {\bibinfo {author} {\bibfnamefont {G.}~\bibnamefont
  {Cognola}}, \bibinfo {author} {\bibfnamefont {M.}~\bibnamefont {Rinaldi}}, \
  and\ \bibinfo {author} {\bibfnamefont {L.}~\bibnamefont {Vanzo}},\ }\href
  {\doibase 10.3390/e17085145} {\bibfield  {journal} {\bibinfo  {journal}
  {Entropy}\ }\textbf {\bibinfo {volume} {17}},\ \bibinfo {pages} {5145}
  (\bibinfo {year} {2015}{\natexlab{a}})},\ \Eprint
  {http://arxiv.org/abs/1506.07096} {arXiv:1506.07096 [gr-qc]} \BibitemShut
  {NoStop}%
\bibitem [{\citenamefont {Cognola}\ \emph
  {et~al.}(2015{\natexlab{b}})\citenamefont {Cognola}, \citenamefont {Rinaldi},
  \citenamefont {Vanzo},\ and\ \citenamefont {Zerbini}}]{Cognola:2015wqa}%
  \BibitemOpen
  \bibfield  {author} {\bibinfo {author} {\bibfnamefont {G.}~\bibnamefont
  {Cognola}}, \bibinfo {author} {\bibfnamefont {M.}~\bibnamefont {Rinaldi}},
  \bibinfo {author} {\bibfnamefont {L.}~\bibnamefont {Vanzo}}, \ and\ \bibinfo
  {author} {\bibfnamefont {S.}~\bibnamefont {Zerbini}},\ }\href {\doibase
  10.1103/PhysRevD.91.104004} {\bibfield  {journal} {\bibinfo  {journal} {Phys.
  Rev.}\ }\textbf {\bibinfo {volume} {D91}},\ \bibinfo {pages} {104004}
  (\bibinfo {year} {2015}{\natexlab{b}})},\ \Eprint
  {http://arxiv.org/abs/1503.05151} {arXiv:1503.05151 [gr-qc]} \BibitemShut
  {NoStop}%
\bibitem [{\citenamefont {Nariai}(1950)}]{Nariai1}%
  \BibitemOpen
  \bibfield  {author} {\bibinfo {author} {\bibfnamefont {H.}~\bibnamefont
  {Nariai}},\ }\href@noop {} {\bibfield  {journal} {\bibinfo  {journal} {Sci.
  Rep. Tohoku Univ.}\ }\textbf {\bibinfo {volume} {34}},\ \bibinfo {pages}
  {160} (\bibinfo {year} {1950})}\BibitemShut {NoStop}%
\bibitem [{\citenamefont {Nariai}(1951)}]{Nariai2}%
  \BibitemOpen
  \bibfield  {author} {\bibinfo {author} {\bibfnamefont {H.}~\bibnamefont
  {Nariai}},\ }\href@noop {} {\bibfield  {journal} {\bibinfo  {journal} {Sci.
  Rep. Tohoku Univ.}\ }\textbf {\bibinfo {volume} {35}},\ \bibinfo {pages}
  {162} (\bibinfo {year} {1951})}\BibitemShut {NoStop}%
\bibitem [{\citenamefont {Sobreiro}\ \emph
  {et~al.}(2012{\natexlab{a}})\citenamefont {Sobreiro}, \citenamefont {Tomaz},\
  and\ \citenamefont {Otoya}}]{Sobreiro:2011hb}%
  \BibitemOpen
  \bibfield  {author} {\bibinfo {author} {\bibfnamefont {R.~F.}\ \bibnamefont
  {Sobreiro}}, \bibinfo {author} {\bibfnamefont {A.~A.}\ \bibnamefont {Tomaz}},
  \ and\ \bibinfo {author} {\bibfnamefont {V.~J.~V.}\ \bibnamefont {Otoya}},\
  }\href@noop {} {\bibfield  {journal} {\bibinfo  {journal} {Eur. Phys. Jour.
  C}\ }\textbf {\bibinfo {volume} {72}},\ \bibinfo {pages} {1991} (\bibinfo
  {year} {2012}{\natexlab{a}})}\BibitemShut {NoStop}%
\bibitem [{\citenamefont {Sobreiro}\ \emph
  {et~al.}(2012{\natexlab{b}})\citenamefont {Sobreiro}, \citenamefont {Tomaz},\
  and\ \citenamefont {Otoya}}]{Sobreiro:2012dp}%
  \BibitemOpen
  \bibfield  {author} {\bibinfo {author} {\bibfnamefont {R.~F.}\ \bibnamefont
  {Sobreiro}}, \bibinfo {author} {\bibfnamefont {A.~A.}\ \bibnamefont {Tomaz}},
  \ and\ \bibinfo {author} {\bibfnamefont {V.~J.~V.}\ \bibnamefont {Otoya}},\
  }\href@noop {} {\bibfield  {journal} {\bibinfo  {journal} {PoS}\ }\textbf
  {\bibinfo {volume} {ICMP2012}},\ \bibinfo {pages} {019} (\bibinfo {year}
  {2012}{\natexlab{b}})},\ \Eprint {http://arxiv.org/abs/1210.8446}
  {arXiv:1210.8446 [hep-th]} \BibitemShut {NoStop}%
\bibitem [{\citenamefont {Sobreiro}\ \emph {et~al.}(2013)\citenamefont
  {Sobreiro}, \citenamefont {Tomaz},\ and\ \citenamefont
  {Vasquez~Otoya}}]{Sobreiro:2012iv}%
  \BibitemOpen
  \bibfield  {author} {\bibinfo {author} {\bibfnamefont {R.~F.}\ \bibnamefont
  {Sobreiro}}, \bibinfo {author} {\bibfnamefont {A.~A.}\ \bibnamefont {Tomaz}},
  \ and\ \bibinfo {author} {\bibfnamefont {V.~J.}\ \bibnamefont
  {Vasquez~Otoya}},\ }\href@noop {} {\bibfield  {journal} {\bibinfo  {journal}
  {J. Phys. Conf. Ser.}\ }\textbf {\bibinfo {volume} {453}},\ \bibinfo {pages}
  {012014} (\bibinfo {year} {2013})}\BibitemShut {NoStop}%
\bibitem [{\citenamefont {Sobreiro}\ and\ \citenamefont
  {Tomaz}(2016)}]{Sobreiro:2016fks}%
  \BibitemOpen
  \bibfield  {author} {\bibinfo {author} {\bibfnamefont {R.~F.}\ \bibnamefont
  {Sobreiro}}\ and\ \bibinfo {author} {\bibfnamefont {A.~A.}\ \bibnamefont
  {Tomaz}},\ }\href {\doibase 10.1155/2016/9048263} {\bibfield  {journal}
  {\bibinfo  {journal} {Adv. High Energy Phys.}\ }\textbf {\bibinfo {volume}
  {2016}},\ \bibinfo {pages} {9048263} (\bibinfo {year} {2016})},\ \Eprint
  {http://arxiv.org/abs/1607.00399} {arXiv:1607.00399 [hep-th]} \BibitemShut
  {NoStop}%
\bibitem [{\citenamefont {Sobreiro}\ and\ \citenamefont
  {Otoya}(2011)}]{Sobreiro:2010qf}%
  \BibitemOpen
  \bibfield  {author} {\bibinfo {author} {\bibfnamefont {R.~F.}\ \bibnamefont
  {Sobreiro}}\ and\ \bibinfo {author} {\bibfnamefont {V.~J.~V.}\ \bibnamefont
  {Otoya}},\ }\href@noop {} {\bibfield  {journal} {\bibinfo  {journal} {J.
  Phys. Conf. Ser.}\ }\textbf {\bibinfo {volume} {283}},\ \bibinfo {pages}
  {012032} (\bibinfo {year} {2011})},\ \Eprint {http://arxiv.org/abs/1010.2946}
  {arXiv:1010.2946 [hep-th]} \BibitemShut {NoStop}%
\bibitem [{\citenamefont {{Hawking}}(1987)}]{1987qftq....2..129H}%
  \BibitemOpen
  \bibfield  {author} {\bibinfo {author} {\bibfnamefont {S.~W.}\ \bibnamefont
  {{Hawking}}},\ }in\ \href@noop {} {\emph {\bibinfo {booktitle} {Quantum field
  theory and quantum statistics, Vol. 2, p. 129 - 139}}},\ Vol.~\bibinfo
  {volume} {2}\ (\bibinfo {year} {1987})\ pp.\ \bibinfo {pages}
  {129--139}\BibitemShut {NoStop}%
\bibitem [{\citenamefont {Hawking}\ and\ \citenamefont
  {Hertog}(2002)}]{Hawking:2001yt}%
  \BibitemOpen
  \bibfield  {author} {\bibinfo {author} {\bibfnamefont {S.~W.}\ \bibnamefont
  {Hawking}}\ and\ \bibinfo {author} {\bibfnamefont {T.}~\bibnamefont
  {Hertog}},\ }\href {\doibase 10.1103/PhysRevD.65.103515} {\bibfield
  {journal} {\bibinfo  {journal} {Phys. Rev.}\ }\textbf {\bibinfo {volume}
  {D65}},\ \bibinfo {pages} {103515} (\bibinfo {year} {2002})},\ \Eprint
  {http://arxiv.org/abs/hep-th/0107088} {arXiv:hep-th/0107088 [hep-th]}
  \BibitemShut {NoStop}%
\bibitem [{\citenamefont {Alvarez}\ \emph
  {et~al.}(2017{\natexlab{a}})\citenamefont {Alvarez}, \citenamefont {Anero},\
  and\ \citenamefont {Gonzalez-Martin}}]{Alvarez:2017spt}%
  \BibitemOpen
  \bibfield  {author} {\bibinfo {author} {\bibfnamefont {E.}~\bibnamefont
  {Alvarez}}, \bibinfo {author} {\bibfnamefont {J.}~\bibnamefont {Anero}}, \
  and\ \bibinfo {author} {\bibfnamefont {S.}~\bibnamefont {Gonzalez-Martin}},\
  }\href {\doibase 10.1088/1475-7516/2017/10/008} {\bibfield  {journal}
  {\bibinfo  {journal} {JCAP}\ }\textbf {\bibinfo {volume} {1710}},\ \bibinfo
  {pages} {008} (\bibinfo {year} {2017}{\natexlab{a}})},\ \Eprint
  {http://arxiv.org/abs/1703.07993} {arXiv:1703.07993 [hep-th]} \BibitemShut
  {NoStop}%
\bibitem [{\citenamefont {Alvarez}\ \emph
  {et~al.}(2017{\natexlab{b}})\citenamefont {Alvarez}, \citenamefont {Anero},
  \citenamefont {Gonzalez-Martin},\ and\ \citenamefont
  {Santos-Garcia}}]{Alvarez:2017ayn}%
  \BibitemOpen
  \bibfield  {author} {\bibinfo {author} {\bibfnamefont {E.}~\bibnamefont
  {Alvarez}}, \bibinfo {author} {\bibfnamefont {J.}~\bibnamefont {Anero}},
  \bibinfo {author} {\bibfnamefont {S.}~\bibnamefont {Gonzalez-Martin}}, \ and\
  \bibinfo {author} {\bibfnamefont {R.}~\bibnamefont {Santos-Garcia}},\ }in\
  \href {http://inspirehep.net/record/1628837/files/arXiv:1710.01764.pdf}
  {\emph {\bibinfo {booktitle} {{18th Lomonosov Conference on Elementary
  Particle Physics Moscow, Russia, August 24-30, 2017}}}}\ (\bibinfo {year}
  {2017})\ \Eprint {http://arxiv.org/abs/1710.01764} {arXiv:1710.01764
  [hep-th]} \BibitemShut {NoStop}%
\bibitem [{\citenamefont {Alvarez}\ \emph {et~al.}(2018)\citenamefont
  {Alvarez}, \citenamefont {Anero}, \citenamefont {Gonzalez-Martin},\ and\
  \citenamefont {Santos-Garcia}}]{Alvarez:2018lrg}%
  \BibitemOpen
  \bibfield  {author} {\bibinfo {author} {\bibfnamefont {E.}~\bibnamefont
  {Alvarez}}, \bibinfo {author} {\bibfnamefont {J.}~\bibnamefont {Anero}},
  \bibinfo {author} {\bibfnamefont {S.}~\bibnamefont {Gonzalez-Martin}}, \ and\
  \bibinfo {author} {\bibfnamefont {R.}~\bibnamefont {Santos-Garcia}},\
  }\href@noop {} {\  (\bibinfo {year} {2018})},\ \Eprint
  {http://arxiv.org/abs/1802.05922} {arXiv:1802.05922 [hep-th]} \BibitemShut
  {NoStop}%
\bibitem [{\citenamefont {Abbott}\ \emph {et~al.}(2016)\citenamefont {Abbott}
  \emph {et~al.}}]{Abbott:2016blz}%
  \BibitemOpen
  \bibfield  {author} {\bibinfo {author} {\bibfnamefont {B.~P.}\ \bibnamefont
  {Abbott}} \emph {et~al.} (\bibinfo {collaboration} {Virgo, LIGO
  Scientific}),\ }\href {\doibase 10.1103/PhysRevLett.116.061102} {\bibfield
  {journal} {\bibinfo  {journal} {Phys. Rev. Lett.}\ }\textbf {\bibinfo
  {volume} {116}},\ \bibinfo {pages} {061102} (\bibinfo {year} {2016})},\
  \Eprint {http://arxiv.org/abs/1602.03837} {arXiv:1602.03837 [gr-qc]}
  \BibitemShut {NoStop}%
\bibitem [{\citenamefont {da~Rocha}\ \emph {et~al.}(2017)\citenamefont
  {da~Rocha}, \citenamefont {Sobreiro},\ and\ \citenamefont
  {Tomaz}}]{daRocha:2017tiz}%
  \BibitemOpen
  \bibfield  {author} {\bibinfo {author} {\bibfnamefont {R.}~\bibnamefont
  {da~Rocha}}, \bibinfo {author} {\bibfnamefont {R.~F.}\ \bibnamefont
  {Sobreiro}}, \ and\ \bibinfo {author} {\bibfnamefont {A.~A.}\ \bibnamefont
  {Tomaz}},\ }\href {\doibase 10.1016/j.physletb.2017.11.009} {\bibfield
  {journal} {\bibinfo  {journal} {Phys. Lett.}\ }\textbf {\bibinfo {volume}
  {B775}},\ \bibinfo {pages} {277} (\bibinfo {year} {2017})},\ \Eprint
  {http://arxiv.org/abs/1705.04877} {arXiv:1705.04877 [gr-qc]} \BibitemShut
  {NoStop}%
\bibitem [{\citenamefont {Assimos}\ \emph {et~al.}(2013)\citenamefont
  {Assimos}, \citenamefont {Pereira}, \citenamefont {Santos}, \citenamefont
  {Sobreiro}, \citenamefont {Tomaz},\ and\ \citenamefont
  {Otoya}}]{Assimos:2013eua}%
  \BibitemOpen
  \bibfield  {author} {\bibinfo {author} {\bibfnamefont {T.~S.}\ \bibnamefont
  {Assimos}}, \bibinfo {author} {\bibfnamefont {A.~D.}\ \bibnamefont
  {Pereira}}, \bibinfo {author} {\bibfnamefont {T.~R.~S.}\ \bibnamefont
  {Santos}}, \bibinfo {author} {\bibfnamefont {R.~F.}\ \bibnamefont
  {Sobreiro}}, \bibinfo {author} {\bibfnamefont {A.~A.}\ \bibnamefont {Tomaz}},
  \ and\ \bibinfo {author} {\bibfnamefont {V.~J.~V.}\ \bibnamefont {Otoya}},\
  }\href@noop {} {\bibfield  {journal} {\bibinfo  {journal} {To appear in the
  International Journal of Modern Physics}\ } (\bibinfo {year} {2013})},\
  \Eprint {http://arxiv.org/abs/1305.1468} {arXiv:1305.1468 [hep-th]}
  \BibitemShut {NoStop}%
\bibitem [{\citenamefont {Gibbons}\ and\ \citenamefont
  {Hawking}(1977)}]{Gibbons:1977mu}%
  \BibitemOpen
  \bibfield  {author} {\bibinfo {author} {\bibfnamefont {G.~W.}\ \bibnamefont
  {Gibbons}}\ and\ \bibinfo {author} {\bibfnamefont {S.~W.}\ \bibnamefont
  {Hawking}},\ }\href@noop {} {\bibfield  {journal} {\bibinfo  {journal} {Phys.
  Rev.}\ }\textbf {\bibinfo {volume} {D15}},\ \bibinfo {pages} {2738} (\bibinfo
  {year} {1977})}\BibitemShut {NoStop}%
\bibitem [{\citenamefont {Cardoso}\ and\ \citenamefont
  {Lemos}(2003)}]{Cardoso:2003sw}%
  \BibitemOpen
  \bibfield  {author} {\bibinfo {author} {\bibfnamefont {V.}~\bibnamefont
  {Cardoso}}\ and\ \bibinfo {author} {\bibfnamefont {J.~P.~S.}\ \bibnamefont
  {Lemos}},\ }\href {\doibase 10.1103/PhysRevD.67.084020} {\bibfield  {journal}
  {\bibinfo  {journal} {Phys. Rev.}\ }\textbf {\bibinfo {volume} {D67}},\
  \bibinfo {pages} {084020} (\bibinfo {year} {2003})},\ \Eprint
  {http://arxiv.org/abs/gr-qc/0301078} {arXiv:gr-qc/0301078 [gr-qc]}
  \BibitemShut {NoStop}%
\bibitem [{\citenamefont {Faraoni}(2015)}]{Faraoni:2015ula}%
  \BibitemOpen
  \bibfield  {author} {\bibinfo {author} {\bibfnamefont {V.}~\bibnamefont
  {Faraoni}},\ }\href {\doibase 10.1007/978-3-319-19240-6} {\bibfield
  {journal} {\bibinfo  {journal} {Lect. Notes Phys.}\ }\textbf {\bibinfo
  {volume} {907}},\ \bibinfo {pages} {pp.1} (\bibinfo {year}
  {2015})}\BibitemShut {NoStop}%
\bibitem [{\citenamefont {Stanley}\ and\ \citenamefont
  {Weisstein}()}]{Weisstein}%
  \BibitemOpen
  \bibfield  {author} {\bibinfo {author} {\bibfnamefont {R.}~\bibnamefont
  {Stanley}}\ and\ \bibinfo {author} {\bibfnamefont {E.~W.}\ \bibnamefont
  {Weisstein}},\ }\href {http://mathworld.wolfram.com/CatalanNumber.html}
  {\enquote {\bibinfo {title} {{C}atalan {N}umbers -- {F}rom {M}ath{W}orld --
  {A} {W}olfram {W}eb {R}esource},}\ }\BibitemShut {NoStop}%
\bibitem [{\citenamefont {Bhattacharya}\ and\ \citenamefont
  {Lahiri}(2010)}]{Bhattacharya:2010vr}%
  \BibitemOpen
  \bibfield  {author} {\bibinfo {author} {\bibfnamefont {S.}~\bibnamefont
  {Bhattacharya}}\ and\ \bibinfo {author} {\bibfnamefont {A.}~\bibnamefont
  {Lahiri}},\ }\href {\doibase 10.1088/0264-9381/27/16/165015} {\bibfield
  {journal} {\bibinfo  {journal} {Class. Quant. Grav.}\ }\textbf {\bibinfo
  {volume} {27}},\ \bibinfo {pages} {165015} (\bibinfo {year} {2010})},\
  \Eprint {http://arxiv.org/abs/1001.1162} {arXiv:1001.1162 [gr-qc]}
  \BibitemShut {NoStop}%
\bibitem [{\citenamefont {Wald}(1984)}]{Wald:1984rg}%
  \BibitemOpen
  \bibfield  {author} {\bibinfo {author} {\bibfnamefont {R.~M.}\ \bibnamefont
  {Wald}},\ }\href@noop {} {\emph {\bibinfo {title} {{General Relativity}}}}\
  (\bibinfo {year} {1984})\BibitemShut {NoStop}%
\bibitem [{\citenamefont {Silveira}\ \emph {et~al.}()\citenamefont {Silveira},
  \citenamefont {Sobreiro},\ and\ \citenamefont {Tomaz}}]{Silveira:2017xxx}%
  \BibitemOpen
  \bibfield  {author} {\bibinfo {author} {\bibfnamefont {F.~A.}\ \bibnamefont
  {Silveira}}, \bibinfo {author} {\bibfnamefont {R.~F.}\ \bibnamefont
  {Sobreiro}}, \ and\ \bibinfo {author} {\bibfnamefont {A.~A.}\ \bibnamefont
  {Tomaz}},\ }\href@noop {} {\ }\BibitemShut {NoStop}%
\bibitem [{\citenamefont {Frolov}\ and\ \citenamefont
  {Novikov}(1998)}]{Frolov:1998wf}%
  \BibitemOpen
  \bibinfo {editor} {\bibfnamefont {V.~P.}\ \bibnamefont {Frolov}}\ and\
  \bibinfo {editor} {\bibfnamefont {I.~D.}\ \bibnamefont {Novikov}},\ eds.,\
  \href@noop {} {\emph {\bibinfo {title} {{Black hole physics: Basic concepts
  and new developments}}}}\ (\bibinfo {year} {1998})\BibitemShut {NoStop}%
\bibitem [{\citenamefont {Hawking}\ and\ \citenamefont
  {Ellis}(2011)}]{Hawking:1973uf}%
  \BibitemOpen
  \bibfield  {author} {\bibinfo {author} {\bibfnamefont {S.~W.}\ \bibnamefont
  {Hawking}}\ and\ \bibinfo {author} {\bibfnamefont {G.~F.~R.}\ \bibnamefont
  {Ellis}},\ }\href@noop {} {\emph {\bibinfo {title} {{The Large Scale
  Structure of Space-Time}}}},\ Cambridge Monographs on Mathematical Physics\
  (\bibinfo  {publisher} {Cambridge University Press},\ \bibinfo {year}
  {2011})\BibitemShut {NoStop}%
\bibitem [{\citenamefont {Bronnikov}\ and\ \citenamefont
  {Rubin}(2012)}]{Bronnikov:2012wsj}%
  \BibitemOpen
  \bibfield  {author} {\bibinfo {author} {\bibfnamefont {K.~A.}\ \bibnamefont
  {Bronnikov}}\ and\ \bibinfo {author} {\bibfnamefont {S.~G.}\ \bibnamefont
  {Rubin}},\ }\href@noop {} {\emph {\bibinfo {title} {{Black Holes, Cosmology
  and Extra Dimensions}}}}\ (\bibinfo  {publisher} {WSP},\ \bibinfo {year}
  {2012})\BibitemShut {NoStop}%
\bibitem [{\citenamefont {Frolov}\ and\ \citenamefont
  {Vilkovisky}(1981)}]{Frolov:1981mz}%
  \BibitemOpen
  \bibfield  {author} {\bibinfo {author} {\bibfnamefont {V.~P.}\ \bibnamefont
  {Frolov}}\ and\ \bibinfo {author} {\bibfnamefont {G.~A.}\ \bibnamefont
  {Vilkovisky}},\ }\href@noop {} {\bibfield  {journal} {\bibinfo  {journal}
  {Phys. Lett.}\ }\textbf {\bibinfo {volume} {B106}},\ \bibinfo {pages} {307}
  (\bibinfo {year} {1981})}\BibitemShut {NoStop}%
\bibitem [{\citenamefont {Hawking}(1975)}]{Hawking:1974sw}%
  \BibitemOpen
  \bibfield  {author} {\bibinfo {author} {\bibfnamefont {S.~W.}\ \bibnamefont
  {Hawking}},\ }\bibfield  {booktitle} {\emph {\bibinfo {booktitle} {{In
  *Gibbons, G.W. (ed.), Hawking, S.W. (ed.): Euclidean quantum gravity*
  167-188}}},\ }\href@noop {} {\bibfield  {journal} {\bibinfo  {journal}
  {Commun. Math. Phys.}\ }\textbf {\bibinfo {volume} {43}},\ \bibinfo {pages}
  {199} (\bibinfo {year} {1975})}\BibitemShut {NoStop}%
\bibitem [{\citenamefont {Kothawala}\ and\ \citenamefont
  {Padmanabhan}(2009)}]{Kothawala:2009kc}%
  \BibitemOpen
  \bibfield  {author} {\bibinfo {author} {\bibfnamefont {D.}~\bibnamefont
  {Kothawala}}\ and\ \bibinfo {author} {\bibfnamefont {T.}~\bibnamefont
  {Padmanabhan}},\ }\href {\doibase 10.1103/PhysRevD.79.104020} {\bibfield
  {journal} {\bibinfo  {journal} {Phys. Rev.}\ }\textbf {\bibinfo {volume}
  {D79}},\ \bibinfo {pages} {104020} (\bibinfo {year} {2009})},\ \Eprint
  {http://arxiv.org/abs/0904.0215} {arXiv:0904.0215 [gr-qc]} \BibitemShut
  {NoStop}%
\bibitem [{\citenamefont {Bamba}\ and\ \citenamefont
  {Geng}(2010)}]{Bamba:2010kf}%
  \BibitemOpen
  \bibfield  {author} {\bibinfo {author} {\bibfnamefont {K.}~\bibnamefont
  {Bamba}}\ and\ \bibinfo {author} {\bibfnamefont {C.-Q.}\ \bibnamefont
  {Geng}},\ }\href {\doibase 10.1088/1475-7516/2010/06/014} {\bibfield
  {journal} {\bibinfo  {journal} {JCAP}\ }\textbf {\bibinfo {volume} {1006}},\
  \bibinfo {pages} {014} (\bibinfo {year} {2010})},\ \Eprint
  {http://arxiv.org/abs/1005.5234} {arXiv:1005.5234 [gr-qc]} \BibitemShut
  {NoStop}%
\bibitem [{\citenamefont {Exirifard}\ and\ \citenamefont
  {Sheikh-Jabbari}(2008)}]{Exirifard:2007da}%
  \BibitemOpen
  \bibfield  {author} {\bibinfo {author} {\bibfnamefont {Q.}~\bibnamefont
  {Exirifard}}\ and\ \bibinfo {author} {\bibfnamefont {M.~M.}\ \bibnamefont
  {Sheikh-Jabbari}},\ }\href {\doibase 10.1016/j.physletb.2008.02.012}
  {\bibfield  {journal} {\bibinfo  {journal} {Phys. Lett.}\ }\textbf {\bibinfo
  {volume} {B661}},\ \bibinfo {pages} {158} (\bibinfo {year} {2008})},\ \Eprint
  {http://arxiv.org/abs/0705.1879} {arXiv:0705.1879 [hep-th]} \BibitemShut
  {NoStop}%
\bibitem [{\citenamefont {Borunda}\ \emph {et~al.}(2008)\citenamefont
  {Borunda}, \citenamefont {Janssen},\ and\ \citenamefont
  {Bastero-Gil}}]{Borunda:2008kf}%
  \BibitemOpen
  \bibfield  {author} {\bibinfo {author} {\bibfnamefont {M.}~\bibnamefont
  {Borunda}}, \bibinfo {author} {\bibfnamefont {B.}~\bibnamefont {Janssen}}, \
  and\ \bibinfo {author} {\bibfnamefont {M.}~\bibnamefont {Bastero-Gil}},\
  }\href {\doibase 10.1088/1475-7516/2008/11/008} {\bibfield  {journal}
  {\bibinfo  {journal} {JCAP}\ }\textbf {\bibinfo {volume} {0811}},\ \bibinfo
  {pages} {008} (\bibinfo {year} {2008})},\ \Eprint
  {http://arxiv.org/abs/0804.4440} {arXiv:0804.4440 [hep-th]} \BibitemShut
  {NoStop}%
\bibitem [{\citenamefont {Capozziello}\ \emph {et~al.}(2011)\citenamefont
  {Capozziello}, \citenamefont {Darabi},\ and\ \citenamefont
  {Vernieri}}]{Capozziello:2010ih}%
  \BibitemOpen
  \bibfield  {author} {\bibinfo {author} {\bibfnamefont {S.}~\bibnamefont
  {Capozziello}}, \bibinfo {author} {\bibfnamefont {F.}~\bibnamefont {Darabi}},
  \ and\ \bibinfo {author} {\bibfnamefont {D.}~\bibnamefont {Vernieri}},\
  }\href {\doibase 10.1142/S021773231103458X} {\bibfield  {journal} {\bibinfo
  {journal} {Mod. Phys. Lett.}\ }\textbf {\bibinfo {volume} {A26}},\ \bibinfo
  {pages} {65} (\bibinfo {year} {2011})},\ \Eprint
  {http://arxiv.org/abs/1006.0454} {arXiv:1006.0454 [gr-qc]} \BibitemShut
  {NoStop}%
\bibitem [{\citenamefont {Dafermos}\ \emph {et~al.}(2016)\citenamefont
  {Dafermos}, \citenamefont {Holzegel},\ and\ \citenamefont
  {Rodnianski}}]{Dafermos:2016uzj}%
  \BibitemOpen
  \bibfield  {author} {\bibinfo {author} {\bibfnamefont {M.}~\bibnamefont
  {Dafermos}}, \bibinfo {author} {\bibfnamefont {G.}~\bibnamefont {Holzegel}},
  \ and\ \bibinfo {author} {\bibfnamefont {I.}~\bibnamefont {Rodnianski}},\
  }\href@noop {} {\  (\bibinfo {year} {2016})},\ \Eprint
  {http://arxiv.org/abs/1601.06467} {arXiv:1601.06467 [gr-qc]} \BibitemShut
  {NoStop}%
\bibitem [{\citenamefont {Chaverra}\ \emph {et~al.}(2013)\citenamefont
  {Chaverra}, \citenamefont {Ortiz},\ and\ \citenamefont
  {Sarbach}}]{Chaverra:2012bh}%
  \BibitemOpen
  \bibfield  {author} {\bibinfo {author} {\bibfnamefont {E.}~\bibnamefont
  {Chaverra}}, \bibinfo {author} {\bibfnamefont {N.}~\bibnamefont {Ortiz}}, \
  and\ \bibinfo {author} {\bibfnamefont {O.}~\bibnamefont {Sarbach}},\ }\href
  {\doibase 10.1103/PhysRevD.87.044015} {\bibfield  {journal} {\bibinfo
  {journal} {Phys. Rev.}\ }\textbf {\bibinfo {volume} {D87}},\ \bibinfo {pages}
  {044015} (\bibinfo {year} {2013})},\ \Eprint {http://arxiv.org/abs/1209.3731}
  {arXiv:1209.3731 [gr-qc]} \BibitemShut {NoStop}%
\bibitem [{\citenamefont {Arfken}\ and\ \citenamefont
  {Weber}(2008)}]{arfken2008mathematical}%
  \BibitemOpen
  \bibfield  {author} {\bibinfo {author} {\bibfnamefont {G.}~\bibnamefont
  {Arfken}}\ and\ \bibinfo {author} {\bibfnamefont {H.}~\bibnamefont {Weber}},\
  }\href {https://books.google.com.br/books?id=f3aCnXWV1CcC} {\emph {\bibinfo
  {title} {Mathematical methods for physicists}}}\ (\bibinfo  {publisher}
  {Elsevier Acad. Press},\ \bibinfo {year} {2008})\BibitemShut {NoStop}%
\bibitem [{\citenamefont {Carroll}(2004)}]{Carroll:2004st}%
  \BibitemOpen
  \bibfield  {author} {\bibinfo {author} {\bibfnamefont {S.~M.}\ \bibnamefont
  {Carroll}},\ }\href@noop {} {\emph {\bibinfo {title} {{Spacetime and
  geometry: An introduction to general relativity}}}}\ (\bibinfo  {publisher}
  {San Francisco, USA: Addison-Wesley (2004) 513p},\ \bibinfo {year}
  {2004})\BibitemShut {NoStop}%
\bibitem [{\citenamefont {Ryder}(2009)}]{Ryder:2009zz}%
  \BibitemOpen
  \bibfield  {author} {\bibinfo {author} {\bibfnamefont {L.}~\bibnamefont
  {Ryder}},\ }\href@noop {} {\emph {\bibinfo {title} {{Introduction to general
  relativity}}}}\ (\bibinfo {year} {2009})\BibitemShut {NoStop}%
\bibitem [{\citenamefont {Hartle}(2003)}]{Hartle:2003yu}%
  \BibitemOpen
  \bibfield  {author} {\bibinfo {author} {\bibfnamefont {J.~B.}\ \bibnamefont
  {Hartle}},\ }\href@noop {} {\emph {\bibinfo {title} {{An introduction to
  Einstein's general relativity}}}}\ (\bibinfo  {publisher} {San Francisco,
  USA: Addison-Wesley (2003) 582 p},\ \bibinfo {year} {2003})\BibitemShut
  {NoStop}%
\bibitem [{\citenamefont {{Misner, Charles W. and Thorne, Kip S. and Wheeler,
  John A.}}(1973)}]{Misner:1974qy}%
  \BibitemOpen
  \bibfield  {author} {\bibinfo {author} {\bibnamefont {{Misner, Charles W. and
  Thorne, Kip S. and Wheeler, John A.}}},\ }\href@noop {} {\emph {\bibinfo
  {title} {{Gravitation}}}}\ (\bibinfo  {publisher} {W. H. Freeman},\ \bibinfo
  {address} {San Francisco},\ \bibinfo {year} {1973})\BibitemShut {NoStop}%
\end{thebibliography}%

\end{document}